\let\includefigures=\iftrue
%
\let\useblackboard=\iftrue
%
%
\newfam\black
\input harvmac
\noblackbox
\includefigures
\message{If you do not have epsf.tex (to include figures),}
\message{change the option at the top of the tex file.}
\input epsf
\def\figin{\epsfcheck\figin}\def\figins{\epsfcheck\figins}
\def\epsfcheck{\ifx\epsfbox\UnDeFiNeD
\message{(NO epsf.tex, FIGURES WILL BE IGNORED)}
\gdef\figin##1{\vskip2in}\gdef\figins##1{\hskip.5in}
\else\message{(FIGURES WILL BE INCLUDED)}%
\gdef\figin##1{##1}\gdef\figins##1{##1}\fi}
\def\DefWarn#1{}
\def\figinsert{\goodbreak\midinsert}
\def\ifig#1#2#3{\DefWarn#1\xdef#1{fig.~\the\figno}
\writedef{#1\leftbracket fig.\noexpand~\the\figno}%
\figinsert\figin{\centerline{#3}}\medskip\centerline{\vbox{
\baselineskip12pt\advance\hsize by -1truein
\noindent\footnotefont{\bf Fig.~\the\figno:} #2}}
\bigskip\endinsert\global\advance\figno by1}
\else
\def\ifig#1#2#3{\xdef#1{fig.~\the\figno}
\writedef{#1\leftbracket fig.\noexpand~\the\figno}%
\global\advance\figno by1}
\fi
%

\def\smallfig#1#2#3{\DefWarn#1\xdef#1{fig.~\the\figno}
\writedef{#1\leftbracket fig.\noexpand~\the\figno}%
\figinsert\figin{\centerline{#3}}\medskip\centerline{\vbox{
\baselineskip12pt\advance\hsize by -1truein
\noindent\footnotefont{\bf Fig.~\the\figno:} #2}}
\endinsert\global\advance\figno by1}

\useblackboard
\message{If you do not have msbm (blackboard bold) fonts,}
\message{change the option at the top of the tex file.}
\font\blackboard=msbm10 scaled \magstep1
\font\blackboards=msbm7
\font\blackboardss=msbm5
\textfont\black=\blackboard
\scriptfont\black=\blackboards
\scriptscriptfont\black=\blackboardss

\else

\fi
%


\def\boxit#1{\vbox{\hrule\hbox{\vrule\kern8pt
\vbox{\hbox{\kern8pt}\hbox{\vbox{#1}}\hbox{\kern8pt}}
\kern8pt\vrule}\hrule}}
\def\mathboxit#1{\vbox{\hrule\hbox{\vrule\kern8pt\vbox{\kern8pt
\hbox{$\displaystyle #1$}\kern8pt}\kern8pt\vrule}\hrule}}

\def\subsubsection#1{\bigskip\noindent
{\it #1}}

\def\yboxit#1#2{\vbox{\hrule height #1 \hbox{\vrule width #1
\vbox{#2}\vrule width #1 }\hrule height #1 }}
\def\fillbox#1{\hbox to #1{\vbox to #1{\vfil}\hfil}}
\def\ybox{{\lower 1.3pt \yboxit{0.4pt}{\fillbox{8pt}}\hskip-0.2pt}}
%
%

\def\ep{\epsilon}
\def\bep{\bar\epsilon}

\def\bsig{\sigma \dag}

\def\bp{\partial \dag}
\def\bthet{\bar \theta }
\def\Dbar{\overline{D} }
\def\bgam{\gamma \dag}
\def\bThet{\Theta \dag}
\def\bbet{\beta \dag}


\def\jb{\bar j}
\def\Qbar{Q \dag}

\def\comments#1{}

\def\p{\partial}

\def\half{{1\over 2}}

\def\ket#1{|#1\rangle}

\def\vev#1{\langle{#1}\rangle}

\def\CD{{\cal D}}

\def\CN{{\cal N}}
\def\CO{{\cal O}}

\def\CL{{\cal L}}

\def\CS{{\cal S}}


\def\II{\relax{I\kern-.10em I}}

\def\IZ{\relax\ifmmode\mathchoice
{\hbox{\cmss Z\kern-.4em Z}}{\hbox{\cmss Z\kern-.4em Z}}
{\lower.9pt\hbox{\cmsss Z\kern-.4em Z}}
{\lower1.2pt\hbox{\cmsss Z\kern-.4em Z}}
\else{\cmss Z\kern-.4emZ}\fi}
\def\IR{\relax{\rm I\kern-.18em R}}
\def\IZ{\relax\ifmmode\mathchoice
{\hbox{\cmss Z\kern-.4em Z}}{\hbox{\cmss Z\kern-.4em Z}}
{\lower.9pt\hbox{\cmsss Z\kern-.4em Z}} {\lower1.2pt\hbox{\cmsss
Z\kern-.4em Z}}\else{\cmss Z\kern-.4em Z}\fi}
\font\cmss=cmss10 \font\cmsss=cmss10 at 7pt
\def\IB{\relax{\rm I\kern-.18em B}}
\def\IC{{\relax\hbox{$\inbar\kern-.3em{\rm C}$}}}
\def\ID{\relax{\rm I\kern-.18em D}}
\def\IE{\relax{\rm I\kern-.18em E}}
\def\IF{\relax{\rm I\kern-.18em F}}
\def\IG{\relax\hbox{$\inbar\kern-.3em{\rm G}$}}
\def\IGa{\relax\hbox{${\rm I}\kern-.18em\Gamma$}}
\def\IH{\relax{\rm I\kern-.18em H}}
\def\II{\relax{\rm I\kern-.18em I}}
\def\IK{\relax{\rm I\kern-.18em K}}
\def\IP{\relax{\rm I\kern-.18em P}}

%

\def\jb{{\bar \jmath}}

\def\inbar{\,\vrule height1.5ex width.4pt depth0pt}

\def\p{\partial}

\def\pb{{\bar \p}}

\font\cmss=cmss10 
\def\IR{\relax{\rm I\kern-.18em R}}

%


%

\def\lp10{\ell_p^{10}}
\def\lp11{\ell_p^{11}}
\def\R11{R_{11}}

\def\frac#1#2{{#1 \over #2}}

\def\dS{\partial \Sigma}


\def\jb{\bar j}
\def\Qbar{Q \dag}

\def\comments#1{}

\def\p{\partial}

\def\half{{1\over 2}}

\def\ket#1{|#1\rangle}

\def\vev#1{\langle{#1}\rangle}

\def\CD{{\cal D}}

\def\CN{{\cal N}}
\def\CO{{\cal O}}

\def\CL{{\cal L}}

\def\CS{{\cal S}}


\def\II{\relax{I\kern-.10em I}}

\def\IZ{\relax\ifmmode\mathchoice
{\hbox{\cmss Z\kern-.4em Z}}{\hbox{\cmss Z\kern-.4em Z}}
{\lower.9pt\hbox{\cmsss Z\kern-.4em Z}}
{\lower1.2pt\hbox{\cmsss Z\kern-.4em Z}}
\else{\cmss Z\kern-.4emZ}\fi}
\def\IB{\relax{\rm I\kern-.18em B}}
\def\IC{{\relax\hbox{$\inbar\kern-.3em{\rm C}$}}}
\def\ID{\relax{\rm I\kern-.18em D}}
\def\IE{\relax{\rm I\kern-.18em E}}
\def\IF{\relax{\rm I\kern-.18em F}}
\def\IG{\relax\hbox{$\inbar\kern-.3em{\rm G}$}}
\def\IGa{\relax\hbox{${\rm I}\kern-.18em\Gamma$}}
\def\IH{\relax{\rm I\kern-.18em H}}
\def\II{\relax{\rm I\kern-.18em I}}
\def\IK{\relax{\rm I\kern-.18em K}}
\def\IP{\relax{\rm I\kern-.18em P}}

%

\def\jb{{\bar \jmath}}

\def\inbar{\,\vrule height1.5ex width.4pt depth0pt}

\def\p{\partial}

\def\pb{{\bar \p}}

\font\cmss=cmss10 
\def\IR{\relax{\rm I\kern-.18em R}}

%


%

\def\lp10{\ell_p^{10}}
\def\lp11{\ell_p^{11}}
\def\R11{R_{11}}

\def\frac#1#2{{#1 \over #2}}

\def\dS{\partial \Sigma}
\def\atbdy{|_{\dS}}


\def\uu{^}
\def\ll{_}

\def\b{\beta}

\def\g{\gamma}

\def\pri{^\prime}

\def\e{\epsilon}

\def\dag{^\dagger}

\def\bt{\tilde{\beta}}
\def\wpt{\tilde{\wp}}


\def\ie{{\it i.e.}}
\def\cf{{\it c.f.}}

\def\eg{{\it e.g.}}
\hyphenation{Di-men-sion-al}

\def\np{{\it Nucl. Phys.}}
\def\prl{{\it Phys. Rev. Lett.}}
\def\pr{{\it Phys. Rev.}}

\def\mpl{{\it Mod. Phys. Lett.}}

\def\ijmp{{\it Int. J. Mod. Phys.}}

\lref\lvw{W. Lerche, C. Vafa and N.P. Warner,
``Chiral rings in $\CN=2$ superconformal theories,''
\np\ {\bf B324}\ (1989) 427.}
\lref\wityang{E. Witten, ``On the Landau-Ginzburg description of
N=2 minimal models,'' \ijmp\ {\bf A9} (1994) 4783,
hep-th/9304026.}
\lref\vafa{C. Vafa, ``String vacua and orbifoldized LG models'',
\mpl\ {\bf A4} (1989) 1169\semi
K. Intriligator and C. Vafa, ``Landau-Ginzburg orbifolds'', \np\
{\bf B339} (1990) 95.}

\lref\bagwit{E. Witten and J. Bagger, ``Quantization of Newton's
Constant in Certain Supergravity Theories,'' {\it Phys. Lett.}
{\bf B115} (1982) 202.}
\lref\sft{C. Lazariou, ``String field theory and brane superpotentials,''
hep-th/0107162\semi
A. Tomasiello, ``A-infinity structure and superpotentials,'' hep-th/0107195.}

\lref\sktalk{Talk given by S. Kachru at Strings 2000,
{\tt http://feynman.physics.lsa.umich.edu/cgi-bin/s2ktalk.cgi?kachru}}

\lref\warner{N. Warner, ``Supersymmetry in boundary integrable models,''
\np\ {\bf B450} (1995) 663, hep-th/9506064}

\lref\Fayet{P. Fayet, ``Higgs Model and Supersymmetry,''
{\it Nuovo Cim.} {\bf A31} (1976) 626.}

\lref\phases{E. Witten, ``Phases of N=2 theories in two dimensions,''
\np\ {\bf B403} (1993) 159, hep-th/9301042.}
\lref\agm{P. Aspinwall, B. Greene and D. Morrison,
``Calabi-Yau moduli space, mirror manifolds and space-time topology
change in string theory,'' \np\ {\bf B416} (1994) 414, hep-th/9309097. }
\lref\cylg{B. Greene, C. Vafa and N. Warner,
``Calabi-Yau manifolds and renormalization group flows,''
\np\ {\bf B324} (1989) 371\semi
E. Martinec, ``Criticality, Catastrophes
and Compactifications,'' in Brink, L. (ed.) et al., {\it Physics
and Mathematics of Strings} 383. }
\lref\web{T.-M. Chiang and B.R. Greene, ``Phases of
mirror symmetry,'' Strings '95 proceedings
{\it Future Perspectives in String Theory} (World Scientific, New Jersey)
1996; hep-th/9509049.}

\lref\Gepner{D. Gepner, ``Exactly Solvable String Compactifications
on Manifolds of $SU(N)$ Holonomy,'' Phys. Lett. {\bf B199} (1987) 380.}

\lref\lastnite{S. Govindarajan, T. Jayaraman and T. Sarkar,
``Disc Instantons in Linear Sigma Models,'' hep-th/0108234;
P.~Mayr,
``N = 1 mirror symmetry and open/closed string duality,''
hep-th/0108229.
}

\lref\Mukai{S. Mukai, ``Symplectic structure of the moduli space of
sheaves on an abelian or K3 surface,'' {\it Invent. Math.} {\bf 77}
(1984) 101.}

\lref\denef{F. Denef, ``Supergravity Flows and D-brane Stability,''
{\it JHEP} {\bf 0008} (2000) 050, hep-th/0005049\semi
F. Denef, B. Greene and M. Raugas, ``Split Attractor Flows and the Spectrum
of BPS D-branes on the Quintic,'' {\it JHEP} {\bf 0105} (2001) 012,
hep-th/0101135\semi
F. Denef, ``(Dis)assembling Special Lagrangians,'' hep-th/0107152.}

\lref\diac{
C.~I.~Lazaroiu,
JHEP {\bf 0106}, 052 (2001), 
hep-th/0102122;
``Unitarity, D-brane dynamics and D-brane categories,''
hep-th/0102183; 
D.-E. Diaconescu, ``Enhanced D-brane Categories from String Field
Theory,'' hep-th/0104200;
C.~I.~Lazaroiu,
JHEP {\bf 0106}, 064 (2001), 
hep-th/0105063;
C.~I.~Lazaroiu, R.~Roiban and D.~Vaman,
hep-th/0107063.
}

\lref\horiiqvafa{K. Hori, A. Iqbal and C. Vafa,
``D-Branes and mirror symmetry,'' hep-th/0005247.}
\lref\govinda{S. Govindarajan, T. Jayaraman, and T. Sarkar,
``On D-branes from gauged linear sigma models,'' hep-th/0007075.}
\lref\diacdoug{D.-E. Diaconescu and M.R. Douglas, ``D-branes
on stringy Calabi-Yau manifolds,'' hep-th/0006224.}
\lref\mayr{P. Mayr, ``Phases of
supersymmetric D-branes on Kaehler manifolds and the McKay correspondence,''
hep-th/0010223.}
\lref\hori{K. Hori, ``Linear models of supersymmetric D-branes,''
hep-th/0012179.}
\lref\hell{S. Hellerman and J. McGreevy, ``Linear
sigma model toolshed for D-brane physics,'' hep-th/0104100.}

\lref\moregovinda{
S. Govindarajan, T. Jayaraman,
``D-branes, exceptional sheaves and quivers on Calabi-Yau manifolds,''
\np\ {\bf B600} (2001) 457, hep-th/0010196;
``Boundary fermions, coherent sheaves and D-branes on Calabi-Yau manifolds,''
hep-th/0104126.}

\lref\morehell{S. Hellerman and J. McGreevy, work in progress.}

\lref\dougcat{M. Douglas, ``D-branes, Categories and ${\CN}=1$
Supersymmetry,'' hep-th/0011017.}
\lref\asplaw{P. Aspinwall and A. Lawrence, ``Derived Categories and
Zero-Brane Stability,'' hep-th/0104147.}
\lref\categoryrefs{
P. Seidel and R.P. Thomas,
``Braid group actions on derived categories
of coherent sheaves,'' math.AG/0001043;
P.S. Aspinwall, ``Some navigation
rules for D-brane monodromy,'' hep-th/0102198;
P. Horja, ``Derived category automorphisms
from mirror symmetry,'' math.AG/0103231.}

\lref\joyce{D. Joyce, ``On counting
special lagrangian homology three spheres,''
hep-th/9907013.}
\lref\shamitjohn{S. Kachru and J. McGreevy, ``Supersymmetric
three-cycles and supersymmetry breaking,'' \pr\ {\bf D61} (2000)
026001, hep-th/9908135.}
\lref\kklmone{S. Kachru, S. Katz, A. Lawrence and J. McGreevy, ``Open
String Instantons and Superpotentials,'' \pr\
{\bf D62} (2000) 026001,  hep-th/9912151.}
\lref\kklmtwo{S. Kachru, S. Katz, A. Lawrence and J. McGreevy, ``Mirror
symmetry for open strings,'' hep-th/0006047.}
\lref\agvafa{M. Aganagic and C. Vafa, ``Mirror Symmetry, D-branes
and Counting Holomorphic Discs,'' hep-th/0012041\semi
M. Aganagic, A. Klemm and C. Vafa, ``Disk Instantons, Mirror Symmetry
and the Duality Web,'' hep-th/0105045.}

\lref\bbs{K.~Becker, M.~Becker and A.~Strominger,
``Five-branes, membranes and nonperturbative string theory,"
\np\ {\bf B456} (1995) 130, hep-th/9507158.}
\lref\ooy{H. Ooguri, Y. Oz and Z. Yin, ``D-branes
on Calabi-Yau spaces and their mirrors,''
\np\ {\bf B477} (1996) 407, hep-th/9606112.}
\lref\bdlr{I. Brunner, M.R. Douglas, A. Lawrence and
C. R\"omelsberger, ``D-branes on the Quintic,''
{\it JHEP} {\bf 0008} (2000) 015, hep-th/9906200.}
\lref\morrpless{D. Morrison and M.R. Plesser, ``Summing the
Instantons: Quantum Cohomology and Mirror Symmetry in Toric Varieties,''
{\it Nucl. Phys.} {\bf B440} (1995) 279, hep-th/9412236.}

\lref\muto{T. Muto, ``D-branes at orbifolds and
topology change,'' \np\ {\bf 521}\ (1998) 183,
hep-th/9711090.}
\lref\dflop{B.R. Greene, ``D-brane topology changing transitions,''
\np\ {\bf B525}\ (1998) 284, hep-th/9711124.}
\lref\mrdstab{M.R. Douglas, B. Fiol and C. R\"omelsberger,
``Stability and BPS branes,'' hep-th/0002037.}
\lref\mrdspectra{M.R. Douglas, B. Fiol and C. R\"omelsberger,
``The spectrum of BPS branes on a noncompact Calabi-Yau,''
hep-th/0003263.}

\lref\dk{J. Distler and S. Kachru, ``(0,2) Landau-Ginzburg Theory,''
Nucl. Phys. {\bf B413} (1993) 213, hep-th/9309110.}
\lref\trieste{J. Distler, ``Notes on (0,2) superconformal
field theories,'' Trieste HEP Cosmology 1994:0322-351, hep-th/9502012.}
\lref\kw{S. Kachru and E. Witten, ``Computing the Complete Massless
Spectrum of a Landau-Ginzburg Orbifold,'' hep-th/9307038.}
\lref\evaed{E. Silverstein and E. Witten, ``Criteria
for conformal invariance of $(0,2)$ models,''
\np\ {\bf 444} (1995) 161, hep-th/9503212.}
\lref\sharpe{E. Sharpe, ``K\"ahler cone substructure,''
{\it Adv.Theor.Math.Phys.} {\bf 2} (1999) 1441, hep-th/9810064.}
\lref\erictom{T. Gomez and E. Sharpe, ``D-branes and Scheme Theory,''
hep-th/0008150.}
\lref\peskin{E. Mirabelli and M. Peskin, ``Transmission of Supersymmetry
Breaking from a 4-Dimensional Boundary,'' Phys. Rev. {\bf D58} (1998)
065002, hep-th/9712214.}
\lref\matrixcy{S. Kachru, A. Lawrence and E. Silverstein,
``On the matrix description of Calabi-Yau compactifications,''
\prl\ {\bf 80}\ (1998) 2996, hep-th/9712223.}
\lref\wittmirr{E. Witten, ``Mirror manifolds and
topological field theory,'' in {\it Mirror Symmetry I},
S.-T. Yau (ed.), American Mathematical Society (1998),
hep-th/9112056.}
\lref\boundaryflow{P. Fendley, H. Saleur and
N.P. Warner, ``Exact solution of a massless
scalar field with a relevant boundary interaction,''
\np\ {\bf B430}\ (1994) 577, hep-th/9406125\semi
J.A. Harvey, D. Kutasov and E.J. Martinec,
``On the relevance of tachyons,'' hep-th/0003101.}
\lref\seiberg{
N.~Seiberg,
Phys.\ Lett.\ {\bf B318}, 469 (1993),
hep-ph/9309335.
}
\lref\joesbook{J. Polchinski, {\it String Theory},
two vols., Cambridge (1999).}
\lref\eisenbud{D. Eisenbud, {\it Commutative
Algebra with a View Toward Algebraic Geometry},
Springer (1996).}
\lref\kmmbsft{D. Kutasov, G. Moore and M. Mari\~no,
``Some exact results on tachyon condensation in string field theory,''
{\it JHEP} {\bf 0010} (2000) 045,
hep-th/0009148.
}
\lref\smallinst{
E.~Witten,
``Small Instantons in String Theory,''
{\it Nucl. Phys.} {\bf B460} (1996) 541,
hep-th/9511030.
}
\lref\professorgriff{P. Griffiths and J. Harris,
{\it Principles of Algebraic Geometry}, John Wiley and
Sons (1978) NY.}
\lref\candelas{P. Candelas, X.C. de la Ossa, P.S. Green
and L. Parkes, ``A pair of Calabi-Yau manifolds
as an exactly solvable superconformal field theory,''
\np\ {\bf 359}\ (1991) 21.}

\lref\nepomechie{
R.~I.~Nepomechie,
``The boundary N = 2 supersymmetric sine-Gordon model,''
Phys.\ Lett.\ B {\bf 516}, 376 (2001);
hep-th/0106207,
``Boundary S matrices with N = 2 supersymmetry,''
Phys.\ Lett.\ B {\bf 516}, 161 (2001),
hep-th/0106223.
}

\Title{\vbox{\baselineskip12pt\hbox{hep-th/0109069}
\hbox{SLAC-PUB-8984}\hbox{SU-ITP-00/24}}}
{\vbox{
\centerline{Linear Sigma Models for Open Strings
}}}
\bigskip
\centerline{Simeon Hellerman,
Shamit Kachru, Albion Lawrence
and John McGreevy}
\bigskip
\centerline{{\it Department of Physics, Stanford University,
Stanford, CA 94305}}
\smallskip
\centerline{{\it SLAC Theory Group, MS 81, PO Box 4349,
Stanford, CA 94309}}

\bigskip
\noindent
We formulate and study a class of massive ${\cal N}=2$ supersymmetric
gauge
field theories coupled to boundary degrees of freedom on the strip.
For some values of the parameters, the infrared
limits of these theories can
be interpreted as open string sigma models describing D-branes in
large-radius Calabi-Yau compactifications.  For other values of
the parameters, these theories flow to CFTs describing branes in more exotic,
non-geometric phases of the Calabi-Yau moduli space such as
the Landau-Ginzburg orbifold phase.
Some simple properties of the branes (like large
radius monodromies and spectra of worldvolume excitations)
can be computed in our model.
We also provide simple worldsheet models of
the transitions which occur at loci of marginal stability,
and of Higgs-Coulomb transitions.

\Date{September 2001}

\newsec{Introduction}

The study of D-branes wrapped on supersymmetric
cycles of Calabi-Yau threefolds serves
the dual purpose of providing explicit
supersymmetric ``brane world'' models,
and of providing probes of substringy distances
in compactifications where quantum geometry comes
into its own.
To date most calculations of the open string
spectra and dynamics in this
class of compactifications have been done at
particular points
in moduli space -- large radius limits, Gepner points,
and CFT orbifold points. Yet a host of important issues
require a more global understanding:
the behavior of D-branes under topology-changing transitions \refs{\dflop,\muto},
the physics of D-brane probes of closed-string singularities \matrixcy,
supersymmetry breaking \shamitjohn, vacuum selection,
and the stability of BPS states \refs{\bdlr,\shamitjohn,\mrdstab,\denef,\mrdspectra}.
A description giving even rough features (such as
spectra and singularities) of a large class of
closed-string backgrounds and D-brane configurations
would be of use.

The gauged linear sigma model (GLSM) \phases\
provides such a description for the type II and
heterotic compactifications with only closed strings.
Exact CFT descriptions are available only
at special points in the moduli space. Instead
one constructs
a massive 2d QFT which has the desired CFT
as an infrared fixed point, and with parameters that
can be mapped onto coordinates on the CFT moduli space.
One may then compute
RG-invariant properties (or properties for
which the behaviour along the flow is understood)
at any point in the moduli space of the Calabi-Yau.
Among other things, the GLSM enables one to
study simple topology-changing processes in string theory, and to obtain
a picture of the ``phase'' structure that arises as one varies
closed string moduli \refs{\phases,\agm,\web}.  The
appearance of Landau-Ginzburg orbifolds as the small-radius
limit of certain Calabi-Yau models \cylg\ is transparent in
the GLSM. One may also use these models
to find singular CFT points in the moduli space
\refs{\phases,\evaed}.

We can describe D-branes
in this picture.  Some pieces of this description have been
developed independently in \refs{\horiiqvafa,\govinda},
especially for ``A-type'' branes wrapping special Lagrangian
submanifolds of the CY \refs{\bbs,\ooy}, and for
some simple ``B-type'' branes (branes
wrapping holomorphic cycles).
For B-type branes the physics of
the LG phase is undeveloped,
so the phase structure of
the open string sector is not understood.
But the B-type branes are particularly useful to study
since the superpotentials for open-string fields
are free from worldsheet instanton corrections
\refs{\bdlr,\kklmone,\kklmtwo}.  In addition,
they are complementary to heterotic $(0,2)$ models
and F-theory compactifications, in that the
data specifying the D-brane configuration
consists of sheaves and bundles on a Calabi-Yau
background.\foot{In practice, for space-filling
branes wrapping cycles in a compact CY, we will also
need to add orientifolds in order to cancel the
RR tadpoles. We leave such a description for future
work.  For a computation of the spectrum of
the D-branes at hand, we can imagine that
the branes sit at a point in the non-compact
spatial directions.}

In this work we construct linear sigma models for
a large class of B-type
branes and describe their phase structure.\foot{Another
description of D-branes in different phases,
inspired by the GLSM description, can be found in
\diacdoug.  In \S5.1 we give some results consistent with their 
analysis.}  The boundary
conditions and bundle data are specified by
adding degrees of freedom on the worldsheet boundary;
they provide the
Chan-Paton factors.  The boundary couplings provide
holomorphic data
which specify the D-brane configurations.
We will describe a class of
D-branes which, when viewed as sheaves on the
threefold, arise naturally in $(0,2)$ models \dk.
Our discussion is
complementary to the heterotic $(0,2)$
GLSMs. For example, the locations of
singular CFTs will be different, as we can
anticipate from heterotic/type I duality.
In the long run, we hope this framework will be useful
for studying the variation of the open
string spectrum as we move through the
open and closed-string moduli space, and
for studying in detail the singularity structure of
these theories.

The outline of our
paper is as follows.
In \S2, we write down
massive models for branes wrapping B-type cycles with
various gauge field backgrounds on the brane, and discuss the phase
structure.  We postpone technical details until \S3,
wherein we describe the relevant supermultiplet structure,
boundary conditions, and worldsheet
Lagrangians.
\S4 presents an alternative technique for
branes of finite codimension which is useful
for describing Higgs-Coulomb transitions.
In \S5, we discuss some applications of our models,
including monodromies and marginal stability transitions
in closed string moduli space,
and branch structures in open string moduli space.
In \S6, we develop methods that one can use to compute the
spectrum of light fields on the brane worldvolume,
and apply this technology to a simple example of a brane
on $K3$.
\S7 contains our conclusions.  We have put a number
of technical details in the appendices.
In Appendix A
we review the
supersymmetry transformations of the bulk multiplets.
In Appendix B we review the
superspace formalism and introduce superspace for boundary
degrees of freedom.
In Appendix C
we explain a formula from homological algebra which we will use
in \S4.

The main ideas of this project were presented
by S.K. at the Strings 2000 conference in Ann Arbor, Michigan \sktalk.  In the
intervening (perhaps overly long) writeup period since then,
several papers
which have significant overlap with our construction have appeared
\refs{\mayr, \hori, \moregovinda}.  Generalizations
of the construction were given in \hell.

\newsec{Physical interpretation and phase structure}

We begin by presenting the crux of the construction,
dispensing with technical details until \S3.
In this section we assume some familiarity with linear
models for closed strings \phases.

Our guiding principle
in constructing these boundary LSMs is the B-type $\CN=2$
supersymmetry which we know we must preserve in the infrared \ooy.
Since this algebra is
nearly identical to $(0,2)$ heterotic
supersymmetry, the multiplet and interaction structure
we employ will be for the most part familiar from
studies of heterotic LSMs \refs{\dk, \trieste}.

Our approach to boundary conditions differs
from previous work on this subject.
The virtue of a linear sigma model
is its trivial UV field space.
All of the nonlinearities are encoded
in the action, and the nonlinear
sigma model arises upon RG flow.
We adopt this
philosophy in describing boundary conditions;
we add boundary potentials for bulk fields,
and interactions between bulk and boundary fields.
We can do this
in a manifestly B-type supersymmetry
invariant way,
before imposing boundary conditions.
The resulting
boundary equations of motion of
the bulk fields should
be satisfied
as boundary conditions.
In the absence of boundary
interactions, a scalar
will satisfy the Neumann condition
that the boundary value is free.
To make a scalar with a Dirichlet condition,
$S(\phi)=0$, say, we find a supersymmetric
way to add to the action a potential
on the boundary $\dS$ of the worldsheet:
$$ \int_{\dS} |S|^2.$$
As in \refs{\boundaryflow},
this potential term dominates the boundary
equation of motion in the infrared.

We will first examine extreme limits of the GLSM
parameters, where
a good approximation to the infrared physics arises
from studying the vacua and light fluctuations evident
in the classical worldsheet action.
These
limits include the large-radius CY, and the
``very small radius'' Gepner point,
as identified in \phases\ and reviewed below.
In these limits we study the vacuum manifold and
the spectrum of massless boundary fermions.  The vacuum
manifold will specify the background CY geometry and
the submanifold on which the branes are wrapped; the
spectrum of massless fermions will identify the Chan-Paton
bundle on these branes.

To be concrete, we will study D-branes
on the quintic CY in $\IP^4$ for which the
bulk $(2,2)$ linear sigma model is well-known \phases.
The field content is: one $U(1)$ gauge multiplet,
five chiral multiplets $\phi^i$ with charge $1$, and
one chiral multiplet $p$ of charge $-5$.
These fields are coupled via a quasihomogeneous bulk superpotential,
$W = pG(\phi)$,
of degree $5$ in the $\phi$'s.

\subsec{Boundary fields}

We want to model a D-brane wrapped on
a supersymmetric cycle of
the quintic, with some gauge bundle $V$.
Let the cycle be $\CS = \{ S^A(\phi) = 0, ~ \forall A =1 \dots l \}$,
a transverse complete intersection.
To do this, add the following matter fields at the boundary of the
worldsheet.\foot{We
label supermultiplets by their lowest component.}
We use $l$ boundary Fermi multiplets
$\gamma_A$, with charges $d^A =- {\rm degree} (S^A(\phi))$ (which will
allow us to cut out the
codimension $l$ cycle of the CY), and
$r+1$ boundary Fermi multiplets
$\beta_{a=1\ldots r+1}$,
with charges $n_a$ (states of which
will supply the Chan-Paton factors).
As in $(0,2)$ models, a fermi multiplet
consists of a complex fermion and a complex auxiliary boson.
We denote the auxiliary
partners of $\gamma_A, \beta_a$ by $g_A, b_a$, respectively.
The Fermi superfields satisfy the chiral constraints:
\eqn\fermiconstr{
\eqalign{
	Q \dag \gamma_A &= 0\cr
	Q\dag \beta_a &= 0
,}
}
where $Q\dag$ is one of the B-type supercharges preserved by the
boundary theory (see Appendix A for a definition of the
supersymmetry transformations).
We also use a boundary chiral multiplet $\wp$ with charge
$-m$.  This multiplet
is not familiar from $(0,2)$ supersymmetry; it
consists of a bosonic component $\wp$ and an auxiliary
fermion component $\xi$ (see \S3).
The short boundary multiplets only differ in the
statistics of their lowest component.  They
contain the same number of degrees of freedom and obey
\eqn\commutator{
   \{\beta, \beta \dag \} = 1, ~~~ [\wp, \wp \dag] = 1.  
}

The important interaction term
for these fields is the boundary superpotential:
\eqn\bsuper{
	 \int_{\partial \Sigma}\int d\theta
	\left(\gamma_A S^A(\phi) +
	\wp F^a(\phi) \beta_{a} \right)
}
(boundary superspace is defined in Appendix B).
$S^A$ is a homogenous polynomial in $\phi$ of degree
$d^A$; while
$f^a$ is a homogenous
$(m-n_a)$th degree polynomial in $\phi$.
In components, this amounts to
\eqn\bsuperincomponents{
	 \int_{\dS} \left( g_A S^A(\phi) +
	 b_a \wp F^a(\phi)
	+ \gamma_A \del_i S^A(\phi) \Theta^i
	+ \beta_a \left( \wp \del_i F^a(\phi) + \xi F^a(\phi) \right)
\right).
}
where the $\Theta$s are the superpartners of the bulk $\phi$ fields.

The $U(1)$ symmetry acting only on boundary fields
\eqn\specialsym{
\eqalign{
\beta_a &\mapsto e^{i \alpha} \beta_a \cr
\wp &\mapsto e^{- i \alpha} \wp
}}
preserves the interaction \bsuper.
In spacetime, it acts as the center-of-mass
$U(1)$ symmetry of the D-brane configuration.
We will gauge this symmetry, and project onto
the sector of states of the boundary theory
with unit charge.

\subsec{Review of bulk phase structure}

The IR fixed point governing the bulk quantum field theory will
be the same as in the $\CN = (2,2)$ case.
For the reader's convenience we
review the story \phases\ for the quintic:
this captures many of the essential features for
CY hypersurfaces and complete intersections
in more general toric varieties.
With $\phi_i, p$ the scalars in the
chiral multiplets, and $\sigma$ the complex scalar in the
vector multiplet, the bosonic potential
in the bulk has the form:
\eqn\quinticpot{
U_{kin} = \vert G(\phi)\vert^2 + \vert p\vert^2 \sum_i \vert {\partial G\over
\partial \phi_i}\vert^2 +
(\sum_i \vert \phi_i \vert^2 - 5\vert p \vert^2 - r)^2 +
\vert \sigma \vert^2 (\sum_i \vert \phi_i\vert^2 + 25 \vert p \vert^2)\ ,
}
where the first two terms arise from the superpotential;
the third term arises from the $D^2$ term after integrating out
the auxiliary field $D$; and the final term is related by supersymmetry
to the gauge-covariant kinetic term.

In semiclassical regimes, the CFT is determined
by the vacuum manifold and the fluctuations around it.
At large positive $r$,
the $D^2$ term requires that some of the $\phi$s are
large and nonzero.
The first term
in \quinticpot\ requires $G=0$.  Since some $\phi^i$ are
nonvanishing, not all $\partial_i G$ can vanish for $G$ transverse, so
$p = \sigma = 0$ and their fluctuations are massive.
Since $p = 0$, the D-term equation plus the $U(1)$ gauge symmetry
forces the $\phi$s to live in $\IP^4$.  The
equation $G=0$ forces $\phi$ to live on the quintic hypersurface
in $\IP^4$.
As $r \to \infty$, $r$ can be identified
with the K\"ahler class of the
quintic CY.  Since $r$, $\phi$, and the gauge coupling
$e^2$  are large in the IR, the fields transverse to the
vacuum manifold are very massive.

At large negative $r$, $p$ must be non-zero due to the
$D^2$ term.  Thus $\sigma = G = \del_i G = 0$.
Since $G$ is transverse, $\phi^i = 0$ and
$|p| = \sqrt{-r/5}$.  The $\phi^i$ are not massive but
have a superpotential $G$.  The vev of $p$ breaks the $U(1)$
gauge symmetry to a $\IZ_5$ which rotates $\phi$ by fifth
roots of unity.  Thus the theory in this phase is a
Landau-Ginzburg orbifold.  The limit $r\to -\infty$,
for $G = \sum_i (\phi^i)^5$, is conjectured to be
an exactly solvable CFT \Gepner.

\subsec{The Calabi-Yau phase}

To determine which D-brane configuration
we are making, and identify the GLSM parameters
with moduli of the infrared CFT, we
consider the large-radius CY phase of the LSM.
After integrating out auxiliary bosons,
the potential energy at the boundary of the string
is
\eqn\boundarypot{
	\sum_A |S^A(\phi)|^2 + \sum_a |\wp F^a(\phi)|^2.
}
The infrared theory will describe fluctuations
about the supersymmetric vacuum, in which
this potential will vanish.
We start by setting $S^A(\phi) = 0$.
The fact that the CY coordinates satisfy
this constraint at the boundaries of the string in the IR
indicates that we are describing a D-brane
wrapped on the algebraic cycle $\CS$.

Next, suppose that the $F^a$ are chosen so that
they do not have a simultaneous zero on the quintic.
This forces
$\wp = 0$ (and leaves $\wp$ with only
massive fluctuations).
$F^a$ also gives a mass to a particular linear
combination of the boundary fermions $\beta^a$ through the
nonvanishing mass term:
\eqn\fmassterm{
	-\sqrt{2} \beta_aF^a\xi.
}
Meanwhile, $\gamma_A$ pair with the bulk fermions
normal to $S = 0$ via the interaction
\eqn\gammamassterm{
	-\sqrt{2} \gamma_A \del_i S^A \Theta^i.
}

The massless fermions will transform in some vector bundle
over $\CS$.  This bundle arises exactly as in heterotic
$(0,2)$ models \refs{\phases,\dk}.
The functions $F^a$ are homogenous polynomials of
order $m - n_a$.  If we choose a section
$s_a$ of $\oplus \CO(n_a)$ over $\IP^4$, then $F^a$ will provide a map
to sections of $\CO(m)$ by contraction of indices.
The bundle $\tilde{V}$ over $\IP^4$ is then defined by the following
exact sequence:
\eqn\seq{
	0 \to \tilde{V} \to \bigoplus_{a=1}^{r+1} {\cal O}(n_a)
		\to {\cal O}(m) \to 0
}
$\tilde{V}$ is the kernel of the map given by $F^a$; $V$
is the restriction of this bundle to $\CS$.  The charge-$n_a$
fermions $\beta_a$ are sections of $\CO(n_a)$, restricted to $\CS$.
The massless fermions are in the kernel of $F^a$ and
therefore live in the bundle $V$.

If the massless boundary fermions
transform as sections of $V$, the Hilbert space
of states that they create will transform in
a $2^r$-dimensional reducible representation of the
structure group.  In flat space with a trivial bundle,
in order to select out Chan-Paton
states in the fundamental representation,
we would project onto states with
precisely one fermion excitation.
The analogue in our Calabi-Yau model is to project
onto states which carry the correct
charge ($+1$) under the boundary symmetry \specialsym.  This
projection and its implementation
will be further discussed in
\S3.

At least some of the open string moduli are manifest in this description of
the D-brane.  Changing the $S^{A}(\phi)$
moves the cycle $\CS$ in its moduli space, while perturbing the $F^a(\phi)$
corresponds to moving in the moduli space of bundles $V \to \CS$.

\subsec{The Landau-Ginzburg phase}

When $r$ is large and negative, the bulk theory is a Landau-Ginzburg
orbifold.
The fluctuations of $\phi$
will be governed by the bulk and boundary potentials.
All of the boundary fermions are massless in this phase.

When
$F$ is set to zero there is no mass term for $\wp$ in the
action.  However, the boundary symmetry
projection allows only a finite number of states of the $\wp$
field;
the
target space {\it does not} develop a noncompact branch.
This is explained in greater detail in \S5.2, and we work
out the spectrum of states in an example in \S6.

\subsec{A few words about quantum corrections}

The LSM is most useful in regimes where a
semiclassical expansion is valid.
One can then reliably identify the light excitations
in the 2d field theory, and the corrections obtained by integrating
out bulk massive modes are suppressed by powers of $1/\vert r\vert$.
The massive boundary modes constitute a finite number of degrees
of freedom and therefore their effects are computable.
An argument along the lines of 
\seiberg\ indicates that the boundary 
superpotential is not renormalized.  

Because we have identified good candidates for the
supercharges and global $U(1)$ symmetries in the infrared theory, as
in \refs{\phases,\wityang,\evaed,\kw} certain
quantities $\it can$ be evaluated reliably in the massive theory.
For instance, changing $r$ is a
$Q$-exact operation in the B-model \wittmirr.  Therefore,
chiral operators in the $Q$-cohomology of the B-model should have
$r$-independent properties, which can be studied without loss of generality
in the regimes where the semiclassical expansion is good.
The simplest example for closed strings is the part of
the chiral ring \lvw\ which is visible in the B-model.  The open
string analogue of this is the spectrum of massless fermionic
open string states
stretched between D-branes (or the corresponding spectrum of
boundary-condition
changing operators), which is computable in topological
open string theory.

Similarly, we can calculate the spectrum of massless fermionic
open string states and the spacetime superpotential
which governs them.
This coupling is $r$ independent
and has been computed (through its correspondence with deformation
theory of curves) in various simple geometric
situations in \refs{\bdlr,\kklmone,
\kklmtwo}; these computations are discussed in the framework of open
string field theory in \sft.
It should be possible to set up the calculation of these
amplitudes directly in the LSM; analogous closed-string calculations
in the LSM framework are discussed in e.g. \refs{\morrpless,\evaed}.

There is some $r$-dependent information of interest:
for example, the phase structure, monodromy matrices,
local behavior at marginal stability transitions,
and features of brane-antibrane systems (\cf\ \S5 and \morehell).
Even in the absence of an analog of the closed string half-twisted
model, it seems likely that in semiclassical regimes
other properties of the CFT should be calculable using the linear model.

\newsec{Ingredients and Details}

We now describe our construction in detail.
For concreteness we phrase
our discussion in the context of the quintic CY.

In order to describe strings ending on
D-brane configurations preserving 4d ${\CN=1}$
supersymmetry, the IR fixed point of our massive
theory must have ${\CN=2}$ superconformal
symmetry.  Furthermore, since closed strings
propagating away from the D-branes will see a background
preserving 4d ${\CN=2}$ supersymmetry, the bulk action of the worldsheet
should have $\CN=(2,2)$ superconformal symmetry broken
to $\CN = 2$ superconformal symmetry by the boundary
theory.

We begin with a massive $\CN=2$ theory with four supersymmetries,
half of which are preserved by the worldsheet boundary.
We assume that these flow to the desired IR supersymmetries.
The bulk multiplets have been described in \phases,
and consist of
a vector multiplet $(v_\alpha,\lambda_\pm,\sigma,D)$,
$5$ chiral multiplets of gauge charge 1, $(\phi^i,\psi^i_\pm, F^i)$
with $i=1\ldots 5$,
and a chiral multiplet of charge -5, $(p, \psi^p_\pm, F^p)$.
Here $\phi$ is a complex scalar; $\psi$ a complex fermion;
$F$ a complex auxiliary scalar; $v_\alpha$ a 2d vector field 
with field stregth $v_{+-} = {i \over q} [P_+, P_-]$;
$\lambda$ a complex fermion; $\sigma$ a complex scalar;
and $D$ a real auxiliary field.
Their transformations under
the bulk $\CN=2$ supersymmetry algebra are reviewed
in Appendix A.


We work on the infinite strip parametrized by the time
coordinate $x_0$ and the spatial coordinate $x_1 \in [0, \pi]$.
We study B-type boundary conditions, which
respect the half of the $(2,2)$ supersymmetry transformations
generated by $Q$ and $Q \dag$ \ooy.


\subsec{B-type supersymmetric bulk terms}

By adding boundary terms
it is possible to write full-superspace bulk terms in
a way which is manifestly invariant under
the B-type supersymmetry which we wish to
preserve.
For a
gauge-invariant bulk operator $\CO$,
\eqn\manifest{
\int d^4 \theta \CO ={1 \over 8} Q Q\dag [S, S\dag ] \CO + \del_0 X + \del_1 Y .
}
Here $S$ and $S \dag$ are the supercharges which are {\it broken} by the
boundary theory, and $X$ and $Y$ are gauge invariant
operators.  The time
derivative term is irrelevant for our purposes, while $Y$
gives a contribution to the boundary action.
Thus, by rearranging the
order in which we act with the bulk supercharges, we can render a
full-superspace integral manifestly B-type supersymmetry invariant.

For a chiral multiplet $\phi$ of charge $q$, the B-type supersymmetric
action is as follows (using the transformations given in Appendix A).
Let us define linear combinations
of $\psi_\pm$ using the notation used for the
B-twisted topological sigma model \wittmirr:
\eqn\twistferm{
\eqalign{
	\eta&=\frac{1}{\sqrt{2}}\left(
	\psi_+ +\psi_-\right)\cr
	\Theta&=\frac{1}{\sqrt{2}}\left(\psi_+ -\psi_-
	\right)\ .
}}

The kinetic terms for the $\phi$ multiplets can be written as
\eqn\qqssphiphi{
\eqalign{
 {1 \over 8} Q Q\dag [ S, S\dag ] \phi \phi\dag =&
\del_0 (\dots) + \cr
& {i \over 2}\left( \tilde \nabla_1 \Theta  \eta \dag 
-\eta (\tilde \nabla_1 \Theta )\dag
+ \eta \nabla_0 \eta \dag + \Theta \tilde \nabla_0 \Theta \dag
\right) \cr
&+ {1 \over 4} F F \dag + \tilde \nabla_0 \phi \nabla_0 \phi\dag -
\tilde \nabla_1 \phi (\nabla_1 \phi) \dag
-  {q  \over \sqrt 2} D | \phi|^2
\cr
&+ q \left(
 \half \phi \left(   \left(\lambda_- \dag + \lambda_+\dag \right)\eta\dag
+ \left(\lambda_- \dag - \lambda_+\dag \right) \Theta \dag \right)
+ {\rm h.c.} \right) \cr
& {i q \over 2} \phi \phi\dag \left(\del_+ \sigma\dag + \del_- \sigma\right)
.}}
Here
\eqn\nablas{
\eqalign{ 
	\nabla_0 \phi \equiv 
	{i \over 2} \{Q, Q\dag\} \phi ~~~~~~
	&\nabla_1\phi \equiv 
	{i \over 2} \{Q, S\dag\} \phi \cr
	\tilde\nabla_0\phi \equiv 
	{i \over 2} \{S, S\dag\} \phi ~~~~~~
	&\tilde \nabla_1\phi \equiv 
	{i \over 2} \{Q\dag, S\} \phi.
}}
Note that $\nabla_1 $ is not anti-hermitean.

For the vector multiplet kinetic terms we get
\eqn\qqsssigmasigma{
\eqalign{
{1 \over 8 } Q Q \dag [S, S \dag ] \sigma \sigma \dag &=
\del_0(\dots) -  \del_+ \sigma \dag \del_- \sigma - \half D^2
- {1 \over 4} v_{+-}^2  \cr
&-  i \left( \lambda_+ \del_- \lambda_+ \dag
			+ \lambda_- \del_+ \lambda_- \dag \right) \cr
&+ \half \del_1 \left\{
\sigma \left( - \del_+ \sigma\dag  -
{ 1\over \sqrt 2} \left( i D - {v_{+-} \over \sqrt 2} \right) \right)
+ \sigma \dag \left( \del_- \sigma  +
{ 1\over \sqrt 2} \left( i D + {v_{+-} \over \sqrt 2} \right) \right)
\right.
\cr
& \left. 
+ i \left( \lambda_{+} \dag \lambda_+ - \lambda_- \dag \lambda_- \right)
\right\}.
}}

The bulk Fayet-Iliopoulos (FI) term and the worldsheet theta term
can also be made manifestly
invariant under B-type supersymmetry.  Take the term:
\eqn\bulkFI{
\sqrt 2 Q \Qbar \sigma = D  + {i \over 2} v_{+-} - \sqrt 2 i \del_1 \sigma \ .}
Let
\eqn\complexFI{
	t = i r + {\theta\over 2\pi}\ .
}
Then
\eqn\totalFI{
	(it) \sqrt{2} Q\Qbar\sigma + h.c. = -r D -
	{\sqrt{2}\theta\over 4\pi}
	v_{+-} +
	{\sqrt{2}\theta\over 2\pi}\del_1\left(\sigma + \bsig\right)
	+ \sqrt 2 r i \del_1 \left(\sigma - \bsig \right).
}

The bulk superpotential term, however, cannot be 
written in a manifestly B-type supersymmetry invariant way 
(as far as we can tell).  Following the above strategy, it can be written as:
\eqn\bulksuperp{
	-i Q S W(\phi) +\ h.c. = - \del_i W F^i + 
	\del_i\del_j W \eta^i\Theta^j +\ h.c
}
Acting on this with $\Qbar$ gives
\eqn\superpvary{
	-i \Qbar Q S W \propto \del_1 (Q W)
}
which gives a nonvanishing boundary term.
This is known as the ``Warner problem,'' since it was pointed out
in \warner; we discuss it further in \S3.7 \foot{This issue 
was studied recently in \nepomechie.}.

\subsec{Boundary multiplets}

In this section we will describe
multiplets of B-type supersymmetry which
live on the boundary of the string.

\bigskip
\noindent{\it Boundary vector multiplets}

There are two candidates for a boundary vector multiplet.
There is a real multiplet, with
superspace expansion
\eqn\realmultexpansion{
 s + \sqrt 2 \theta \lambda - \sqrt 2 \bar \theta \lambda \dag +
\theta \bar \theta d 
}
where $s$ is a real scalar, $d$ is a real auxiliary boson, and
$\lambda$ is a complex fermion.  For B-type supersymmetry, this multiplet
only arises in the reduction of the bulk vector multiplet.

We may also define a real supersymmetry singlet.
We shall refer to this as a boundary vector multiplet.
It is just a boundary gauge field $a_0$ which is annihilated by
all supercharges; we will include such a multiplet
to gauge the boundary symmetry \specialsym.

\bigskip
\noindent{\it Fermi multiplets}

We can define boundary Fermi multiplets following
the discussion of $(0,2)$ models in \phases.
They consist of a boundary fermion $\gamma$ and
a boundary auxiliary field $g$.
Suppose $\gamma$ has charge $q$ under a boundary vector multiplet $v$.
The superfield satisfies the chiral constraint
\eqn\boundfermcon{
	Q \dag \gamma = E
}
where $E$ is any boundary chiral boson ($Q\dag E = 0$) with charge $q$
and components $(E,\psi_E)$.
For example, $E$ can be a function of the boundary values
of the bulk chiral fields.

The supersymmetry transformations of this multiplet are:
\eqn\boundfermsusy{
\eqalign{
	Q\gamma = g &~~ Q \dag \gamma = E  \cr
	Qg = 0 &~~ Q \dag g = - 2 i \nabla_0 \gamma + i \xi_E.
}}

\bigskip
\noindent{\it Boundary chiral multiplets}

Boundary chiral multiplets consist of
a complex scalar $\wp$ and a complex fermion
$\xi$.  There is no auxiliary boson.
The superfield satisfies the chiral constraint
\eqn\deformedconstraintonwp
{
 Q \dag \wp = \tau
}
where $(\tau, g_\tau)$ is a fermi multiplet of the same charge as $\wp$
satisfying $Q \dag \tau =0 $.
The supersymmetry transformations are:
\eqn\bchsusy{
\eqalign{
	Q \wp = - i \xi &~~ Q \dag \wp = -i \tau \cr
	Q \xi = 0 &~~ Q \dag \xi = 2 \nabla_0 \wp - g_\tau
}}
We will see that the fermion $\xi$ will in general have no
kinetic terms and will be massive;
it can be integrated out algebraically.

\subsec{Boundary terms in the action}

In this section we describe supersymmetric
actions for boundary fields,
including their couplings to the bulk chiral and vector
multiplets.

\bigskip
\noindent{\it Kinetic terms}

As in \hell\ we use first-order kinetic terms for
short boundary multiplets of both statistics.
Assume $E$ is a holomorphic function of boundary chiral
superfields $\chi^I$.  Then the following kinetic term
is supersymmetric:
\eqn\bfermikinetic{
\eqalign{
	S_{fermi} &= \int dx^0 d^2 \theta \gamma \dag \gamma \cr
	&=
	\int dx^0 \left[ i \left(
	\bgam \nabla_0 \gamma - \nabla_0 \bgam \gamma\right)
	\right.\cr
	&\ \ \ \left. - i \bgam\del_I E(\chi)\xi_\chi ^I - i \gamma\pb_{I}
		E\dag(\chi\dag) \xi_\chi^{I\dagger}
		+ |g|^2 - |E|^2\right]
}}

For a boundary chiral multiplet $\wp$, we add a magnetic field term
\eqn\bchiralkinetic{
\CL = \int d^2 \theta ~ B \wp \dag \wp =
B \left( i  (\nabla\ll 0 \wp) \wp\dag -  i \wp(\nabla\ll 0 \wp\dag)
-  \xi\xi\dag + \tau \tau \dag - i \left( g_\tau \wp \dag - g_\tau \dag \wp \right) 
\right).
}
A second order kinetic term for a boundary
chiral multiplet is less relevant
and therefore flows away in the IR; we omit it
from the outset.
Canonical quantization then gives
\eqn\canonicalreln{
 [ \wp, \wp^\dagger ] = 1/B .
}
The coupling to the magnetic field masses up the fermion $\xi$ and
halves the number of $\wp$ degrees of freedom.  The set of
states
made by $\wp$ is the Hilbert space of a harmonic oscillator.
The number operator $\wp \dag \wp$ for this oscillator is now the generator
of phase rotations of $\wp$.
The $\wp$ multiplet becomes the bosonic analog of a Fermi
multiplet.
Henceforth, we normalize $\wp$ so that $B\equiv 1$.

\bigskip
\noindent{\it Superpotential terms}

The crucial supersymmetry invariant for
our purposes is the boundary superpotential term, which
is an integral over half of the B-type superspace.
We will use terms of this form to specify our brane
configuration and its gauge bundle.
Take a set of Fermi
multiplets $\gamma_A$, such that
$ Q \dag \gamma_A = E_A(\chi)$, 
with $\chi$ representing any scalar multiplet on the boundary 
with $Q \dag \chi^I = \tau^I$.  
Given a collection
of holomorphic functions $S^A(\chi)$ of the scalar multiplets,
the object
\eqn\boundarysuper{
\eqalign{
	S_{super} &= \int dx^0 d\theta
		\sum_A \gamma_A S^A(\chi) + h.c.\cr
	&= \sum_{A}\int dx^0 \left[
		\sum_I 	  g_A S^A 
 + i \p_I S^A \gamma_A \xi_\chi^I \right] + h.c.\ .
}}
is invariant under supersymmetry if 
$$ Q \left( E_A S^A - \gamma_A \del_I S^A \tau^I \right) = 0 .$$
This is easily satisfied if $E_A S^A$ is a constant and 
if $S^A$ is chiral; this will be the case for all
of the examples we consider.
Upon integrating out the auxiliary bosons,
$g_A$, one finds a boundary potential
\eqn\boundaryenergyfromS{
 V = |S^A(\wp)|^2 .
}





\subsec{Boundary conditions}

Next, we work out the implications of the boundary terms
in \S3.1 for the boundary values
of the bulk fields.

Consider first the case with no boundary superpotential.
Then, the variation of the bulk action under variation of
$\phi \dag$ at the boundary is
\eqn\phiboundaryeom{
\int_{\dS} \left( - D_1 \phi \right) \delta \phi \dag.
}
Thus, since we allow arbitrary variations of $\phi \dag$ at the
boundary, we find
\eqn\neumannagain{
	D_1 \phi \atbdy = 0\ .
}

If the bulk action is B-type supersymmetric (\cf\
\S3.1), the boundary conditions on fermions
will be the ones implied by
supersymmetry from the conditions on the bosons.
The boundary equation of
motion for $\Theta\dag$ from \qqssphiphi\ is
\eqn\fermionboundary{ 0 = \int_{\dS} \delta \Theta \dag \eta .
}
Since $\delta \Theta \dag $ is arbitrary,
\eqn\etagone{
	\eta\atbdy = 0
}
and in particular there is no $\eta$ zero mode, as we would
expect from \eg\ Neumann conditions on a superstring
in flat space \joesbook.  This is consistent with
the fact that
\eqn\superbc{ Q \eta \atbdy \sim D_1 \phi \atbdy .}

Next we explain how boundary superpotentials effect
supersymmetric Dirichlet conditions.
Add a boundary superpotential of
the form
\eqn\again{
\int_{\dS} d \theta \gamma S(\phi).
}
Upon addition of such a boundary superpotential, 
the boundary equation of motion for a bulk scalar $\phi^i$ becomes
\eqn\modgenterm{
\eqalign{
	\int dx^0 \delta\phi^i \left(
	\eta_{i\jb}D_1 \phi^{\jb \dagger}
		+ g_A\del_i S^A
		- \del_i E_A E_A\dag
	+ i \bgam_A\p_i\p_I E_A \xi^I
	- i \del_i\del_I S^A\gamma_A \xi^I \right) \cr
  	= \int dx^0 \delta\phi^i \left(
	\eta_{i\jb}D_1 \phi^{\jb \dagger}
		-  \del_i S^A S^{A \dagger}
		-  \del_i E_A E_A\dag
	+{\rm fermions} \right).
}}
(In the second step we have integrated out the boundary
auxiliary bosons.)
In the UV these are some ``rotated'' boundary conditions.
In the IR
the dominant terms come from the bulk and boundary potentials
and the derivative terms in \modgenterm\ can be ignored.
This is as in \boundaryflow.
Fluctuations in directions in field space
$n^i$ with
$$n^i {{\partial S} \over {\partial \phi_{i}}} \neq 0
~~~ {\rm or} ~~~
n^i {{\partial E} \over {\partial \phi_{i}}} \neq 0
$$
will be energetically forbidden; 
they are effectively 
frozen to zero by Dirichlet boundary conditions in the IR.
On the other hand, in field space directions
for which
${\del \over \del \phi_i} S$ and $ {\del \over \del \phi_i} E$
both vanish,
the
boundary equation of motion will impose the Neumann condition
$D_1 \phi^i |_{\dS} = 0$ as usual.
In this manner,
the IR dynamics
yield
effective Dirichlet conditions,
$S(\phi)|_{\dS} = E(\phi)|_{\dS} = 0$, with Neumann conditions
in all other directions.

When we add \again\ the variation
of the action with respect to $\bThet$ is:
\eqn\correctedfermbc{ 0 = \int_{\dS} \delta \bThet_i
\left( \eta_i - \gamma {\del \over \del \phi^i} S(\phi) \right).
}
Thus, the boundary fermion $\gamma$ supplies the
zero mode for $\eta$ normal to the hypersurface
$\{ S(\phi) = 0 \}$.  This zero mode
is required by supersymmetry if the bosonic partner has Dirichlet
conditions.

Boundary equations of motion for the bulk vector multiplet
are as follows.
The variation of $v_0$ gives
\eqn\boundarytermvector{ 0 =
\delta S = \int_{\dS} \delta v_0 \left( -  v_{+-} - \theta + j_0\atbdy \right),
}
where $j_0$ is the gauge current coupling to $v_\alpha$.

The boundary variation of $\sigma, \bsig$ gives
\eqn\sigmaboundaryeom{
\int_{\dS}
\left( \half \del_1 \sigma +
{ 1 \over \sqrt 2} \left( 2 i r + 2 \theta +
{i \over 2} D + { v_{+ -} \over 2 \sqrt 2} \right)
\right) \delta \bsig
- \half \sigma \del_1 \delta \sigma \dag + {\rm h.c.}.
}
We note here that the vanishing of this term is consistent with the
A-type boundary condition
\eqn\bcsforsigma{
 \sigma - e^{i \gamma} \sigma \dag \atbdy = 0
}
with Neumann conditions on the orthogonal combination.  The angle $\gamma$
that this line in the complex $\sigma$ plane makes with the real axis
is unfixed.

Finally, the equations of motion together with the
supersymmetry variations of $\sigma$ require that $\lambda$
satisfy the boundary condition

\eqn\atypeonlambda{
e^{i\gamma} \lambda_{+} + \lambda_{-} \atbdy = 0\ .}


\subsec{Bulk multiplets on the boundary}

In order to couple bulk and boundary multiplets
supersymmetrically, we can decompose the
boundary values of bulk
multiplets into boundary multiplets.

The bulk vector multiplet restricts
to a boundary vector multiplet and a real multiplet.
The combination
\eqn\susysinglet{
	\nabla_0 = {i \over 2} \{Q, Q\dag\}
}
is a singlet under the B-type $\CN=2$ supersymmetry.
It will couple minimally to charged
boundary matter fields according to their
gauge charge under the bulk $U(1)$ symmetry.

The story for bulk chiral multiplets depends
on the boundary conditions.
If $\phi$ does not appear in a boundary superpotential,
then:
\eqn\neumgone{
\eqalign{
   \nabla_1 \phi |_{\partial \Sigma} &= 0 \cr
    \eta |_{\partial \Sigma} &= 0 \cr
    \nabla_1 \Theta |_{\partial \Sigma} &= 0 \ .
}}
The supersymmetry variations
in Appendix A imply that
$\phi$ and $\Theta$ form a boundary chiral multiplet, and
$\eta$ and $F$ form a (trivial) boundary fermi multiplet.
RG flow may induce a magnetic field coupling for the charged bulk
fields at the boundary.
This would make them analogous to the $\wp$ multiplet
described above.
Our considerations will be insensitive to this issue.

If $\phi$ is fixed by a boundary superpotential,
\eqn\dirgone{
\eqalign{
   \nabla_0 \phi|_{\dS} &= 0 \cr
   \Theta |_{\dS} &= 0 \ .
}}
In this case,
the supersymmetry transformations imply that $\eta, F$
form a Fermi multiplet with a deformed
chiral constraint $Q \dag \eta \sim \nabla_1 \phi.$

\subsubsection{A boundary Lagrangian}

Next we write down the complete boundary
interactions in an example.
Take a Fermi multiplet $\beta$ such that
$\Dbar\beta_a = \sqrt{2}\wp E_a(\phi)$,
with superpotential coupling
\eqn\againagain{
 \int_{\del\Sigma} dx^0d\theta \beta_a \wp F^a(\phi)\ ,
}
where $\wp$ is a boundary chiral multiplet such that
$\Dbar\wp = 0$.  Then the kinetic and superpotential
terms are:
\eqn\bact{
\eqalign{
	S_B & = \sum_a \int dx^0 \left[ i \left(
	\beta\dag_a \nabla_0 \beta_a
		- \nabla_0 \beta\dag_a \beta_a \right)
	+ |b_a|^2 - |\wp E_a|^2\right.\cr
	&\ \ \ \ \ + \left( - i \beta\dag_a E_a \xi
		- i \beta\dag_a \wp \del_i E_a \Theta^i + {\rm h.c.} \right)\cr
	&\ \ \ \ \ \left.
	+ \left( i \beta_a F^a \xi
	+ i \beta_a \wp \sum_i \del_i F_a \Theta^i
	+ b_a \wp F^a + {\rm h.c.} \right)\right]\ .
}}

If we integrate out the boundary auxiliary fields,
the boundary Lagrangian becomes:
\eqn\bospot{
\eqalign{
\CL_{\del \Sigma} &=
	- \sum_a \left( 2 |\wp|^2|F^a|^2 + |\wp|^2 |E_a|^2 \right) \cr
& + i  \left(
	\beta_a\dag \nabla_0 \beta_a
		- \nabla_0 \beta\dag_a \beta_a \right)
-  i \left(\nabla_0 \wp\dag \right) \wp + i \wp \dag \nabla_0 \wp \cr
&+ \left( - i \sum_a \beta_a \dag |E_a|^2 \beta_a  
		-i \beta_a \dag \wp \del_i E_a \Theta^i + {\rm h.c.} \right)
	+ \left(  - \sum_a \beta_a\dag  |F^a|^2 \beta_a
	+ i \beta_a \wp \sum_i \del_i F_a \Theta^i
	+ {\rm h.c.} \right) \cr
&+  (i \nabla_1 p )\dag p + \eta_p\dag \Theta_p + {\rm h.c.} \cr
&+ a_0 \left( j_s - 1 \right) \cr
&+ \sigma
\left(- \del_+ \sigma \dag - {1 \over \sqrt 2 } \left( i D - {v_{+-} \over \sqrt 2} \right) \right)
+ \sigma \dag
\left( \del_- \sigma + {1 \over \sqrt 2 } \left( i D + {v_{+-} \over \sqrt 2} \right) \right) \cr
&+ i \left( \lambda_+ \dag \lambda_+ - \lambda_- \dag \lambda_- \right)
+ {1 \over \sqrt 2} \left( \sigma \dag t \dag - \sigma t \right)
- {i \over 4} \left( \sigma - \sigma \dag \right)
\left( \sum _i q_i |\phi_i|^2 \right) .
}}
The boundary term involving the bulk $p$ field will be explained in
\S3.7.

\subsec{Gauging the special symmetry}

As mentioned in \S2,
because the massless boundary
fermions live in a vector bundle $V$, the
Hilbert space on the boundary will transform in a sum
of representations of the structure group of $V$.
The $\beta$s create $2^r$ states.
To find the correct
rank-$r$ Hilbert
space at one end of the string, we make a projection.

To motivate this, consider $r$ coincident branes in flat space.
We can model this by introducing
$r$ free boundary fermi multiplets $\beta_a$.
These fermions acting on the vacuum give
$2^r = 1 + r + \left(^r_2\right) + \cdots$ states.
To obtain an $r$-dimensional space of Chan-Paton
factors, we project onto states with
boundary fermion number one.

Since our boundary fermions interact,
boundary fermion number is no longer
conserved.  However, the boundary fermion number symmetry
is replaced by the global boundary
symmetry \specialsym\ mentioned above.
We will project onto states which have charge $1$
under this symmetry.  This is accomplished by including a
boundary vector, $a_0$, to
act as a Lagrange multiplier

\eqn\bndrylagrange{  \CL_{\rm bdy} \ni  a_0 (j_s - 1)\ ,}
where
\eqn\bndrycurrentdef{
j_s = ~: \bbet\beta: - : \wp\dag \wp :}
is the boundary symmetry current.

\subsec{The Warner problem}

In the presence of a B-type boundary, the
bulk worldsheet superpotential is no longer supersymmetric.
Its $Q\dag$ variation is:
\eqn\warnerterm{
\eqalign{
	Q\dag \int_{\Sigma} \int d^2 \theta ~W
	&=\int dx^0
	\Theta^{\jb\dagger}
	\frac{i}{\sqrt{2}}\left(\eta_{i\jb}F^i -
	2\pb_\jb W\dag \right) \cr
	&\sim
	\int_{\dS} Q W.
}}
This does not vanish even if we impose the equations
of motion for $F$.

There are many options for dealing with this term.
There is a family of boundary terms whose $Q\dag$
variation cancels the Warner term when the auxiliary
fields in the bulk chiral multiplets satisfy their
equations of motion.
For example for the quintic with
$$W = pG(\phi)$$
one can add
\eqn\cancellation{ \Delta \CL_{\rm bdy} = \int d\theta ~ p \eta^{p \dagger} + {\rm h.c.} }
which has
\eqn\warnervariation{
 Q\dag \int d\theta ~ p \eta^{p \dagger }
= \int d\theta ~ p F^{p \dagger } .
}
When the auxiliary field $F^p$ is evaluated on-shell, this gives
\eqn\cancellationmotherearth
{
 Q \dag \Delta \CL_{\rm bdy}
\sim \int d\theta ~ p G(\phi) + {\rm h.c.} 
}
which cancels the Warner variation.
This solution works for any quasihomogeneous bulk
superpotential.

Another approach is to set $W = 0$ as one of the
boundary conditions for the bulk chiral multiplets.
Then the natural fermionic partner of this
condition is precisely $\Theta^i\p_i W = 0$,
and the boundary term \warnerterm\ vanishes.
This is natural in that it
simply forces the boundaries of the worldsheet to
lie in the CY hypersurface.
This is equivalent to adding a
neutral fermi multiplet $M = \mu + \theta m + \cdots$ such that
\eqn\warnerfixtwoferm{
 {\overline D} M = - \frac{1}{\sqrt{2}}
}
and writing a term of the form \boundarysuper\
with $S = W$.  The supersymmetry variation of this
action will
cancel
\warnerterm.  This is the off-shell version of the fix used in
\warner.  The constraint \warnerfixtwoferm\ is the origin of
the inhomogenous terms in the supersymmetry transformation
in \warner.

We believe that both these solutions will
lead to sensible conformal theories.

\subsec{Symmetries of bulk and boundary fields}

\bigskip
\noindent{\it Bulk fields}

We will first review the global symmetries
of the bulk fields \phases.
There are left- and right-moving R-symmetries
acting on the bulk chiral fields.
Under the right-moving
R-symmetry
the supersymmetry current $Q^-$ has charge 1\foot{
The relation between the chiral bulk supercharges
$Q^\pm$ and the B-type supercharges $Q, Q\dag$
is described in the Appendices.};
the Grassman variable $\theta^+$ has charge 1;
and the fields $(\psi_+,F,\sigma,\lambda_-)$ have
charges $(-1,-1,1,1)$.  Under the left-moving R-symmetry,
the supersymmetry current
$Q^+$ has charge 1; the Grassman variable
$\theta^-$ has charge 1; and the fields $(\psi_-,F,\sigma,\lambda_+)$
have charges $(-1,-1,-1,1)$.  These symmetries are
non-anomalous so long as the sum of charges of the
chiral multiplets vanishes.

The bulk superpotential generically breaks the R-symmetry.
However, in absence of the superpotential there was
an additional global symmetry
under which each superfield $\Phi^i$ could be assigned
arbitrary charge $k_i$.
By adding this symmetry to
the R-charge, we find new left and right moving
$R$ symmetries under which $W$ has left- and right- moving
R-charge $1$.  This symmetry is diagonal;
it must be added to both the left- and right-moving R-charges.

For B-type boundary SCFTs, only the sum of the R-charges
is preserved:
\eqn\totalrch{
	R_{tot,i} = R_+ + R_- + 2 k_i\ .
}

\bigskip
\noindent{\it Boundary fields}

Since bulk and boundary fields are coupled,
both will be charged under the conserved R-symmetry.
The R-charges of the boundary fields
can be deduced from the
charges of bulk fields which restrict to the boundary.
One will again have to compose symmetries of the boundary theory
with the naive R-symmetry to make a symmetry which is
respected by the interactions.
There is no anomaly in $(0+1)$ dimensions so
this symmetry will survive quantization.
We will assume that
this $U(1)_R$ flows to the expected R-symmetry of the infrared
${\cal N}=2$ superconformal algebra.

\newsec{Sheafy variables}

Given a description
of a Chan-Paton bundle as the cohomology of an arbitrary sequence of
sums of line bundles, a prescription was given in \hell\
for making a linear sigma model of the type detailed above.
However, \hell\ focused on bundles which have constant fiber dimension
over the entire Calabi-Yau manifold.

One linear model for
branes of finite codimension, \ie\ sheaves with
nonconstant fiber dimension over the CY, was
described in \S2.3.  Simply
add a boundary fermi multiplet $\gamma$ of charge
$-d \equiv - {\rm degree}(S(\phi))$
and add a
boundary superpotential
\eqn\boundarysuperpotential{\int d\theta \gamma S(\phi).}
This produces a boundary energy $\sim |S(\phi)|^2$ which
confines the string boundary to the hypersurface.
This boundary theory would describe a D-brane localized at $S(\phi) = 0$.
In this section we present another formulation of this
model, which is useful because it makes e.g.
the possibility of transitions between Higgs and Coulomb branches in the
D-brane moduli space more transparent.

Introduce a boundary fermi multiplet $\beta$ of
bulk gauge charge $-d$ , and a
gauge neutral boundary chiral multiplet $\wp$, and project as
usual onto the charge one sector of the
boundary symmetry \specialsym.  Instead
of \boundarysuperpotential, add
\eqn\sheafysubcyclesuper{
\int d \theta \wp \beta S(\phi)\ .
}
The resulting vacuum equations set $\wp S(\phi) = 0$.
Away from the locus $\CS \equiv \{S=0\} $ this is
accomplished by setting $\wp = 0$; $\beta$ is
massed up by $\xi$.
Away from the hypersurface, therefore, there
is no way to satisfy the boundary charge projection
on the vacuum manifold.  When $S=0$, on the other hand,
there is no constraint on $\beta$ or $\wp$, but the charge
projection picks out a single Chan-Paton state which
we will describe in detail below.
The resulting D-brane is the sheaf
which is the cohomology of the
sequence
\eqn\skyscraper{ 0 \to \CO (-d) {\buildrel S(\phi) \over\longrightarrow} \CO \to 0 .}
Away from $\CS$ there is no cohomology; over the hypersurface,
the map degenerates and a single cohomology element
appears at adjacent nodes.

This method of building a brane on a subcycle
has also been described
from the spacetime point of view,
\eg\ in \refs{\dougcat,\asplaw,\diac }.
At large volume, adjacent vector bundles in
a sequence have opposite-signed D6-brane charge.
The map between them represents the open string tachyon.
The sequence above describes a D6-brane with D4-brane
charge, and an anti-D6 brane; they annihilate to
leave a D4-brane behind.

To translate between the two descriptions of hypersurfaces
we have given, one can think of $\gamma$ as a
boundary-gauge-symmetry-invariant composite of $\wp$ and $\beta$.
We refer to the description using $\wp$ and $\beta$
as ``sheafy variables'' and employ
it in the next section when
we discuss transitions in open-string moduli space.

The reader may be confused by the fact that in the sheafy
description there seem to
be {\it two} Chan-Paton states localized at the zero set $\CS$,
one created by $\beta$ and one created by $\wp \dag$.  That this
is not the case can be seen by calculating the cohomology of
$Q\dag$ acting on boundary states.  As explained in more detail
in \S6, this yields the number of massless open string states
in the Ramond sector.

Consider an example where there is just a single bulk chiral field $z$
parametrizing a copy of $\IC$.
Take $\CS \equiv \{S(z)=z=0\}$,
\ie\ a D0-brane at the origin,
as described by the sequence \skyscraper.
Finally, introduce an additional
space-filling brane with a single trivial Chan-Paton state,
\ie\ a D2-brane.
We will find the massless strings
stretching between our brane
made by the sequence \skyscraper\ and the D2-brane.
Let us start with a vacuum of the interesting end of the string
which satisfies
\eqn\dzdtvacstate{0 = \Theta \ket{0} = \beta \dag \ket{0} = \wp\ket{0}.}
Imposing the boundary charge projection,
an arbitrary state in the Hilbert space is
\eqn\arbitrary{
\ket{\psi} = f_1(z, z\dag)\beta \ket{0} + f_2(z, z\dag) \beta \Theta \dag \ket{0}
+ f_3(z, z\dag) \wp\dag \ket{0} + f_4(z, z\dag) \wp \dag \Theta \dag \ket{0} .}
Now we want to solve for $\Qbar\ket{\psi} = 0$
modulo $\Qbar$-exact states.  
Reducing to zeromodes, the relevant terms in $\Qbar$ are
\eqn\superchargeforthisexample{
Q \dag = \Theta\dag \overline{\del} + \beta \wp z.
}
Acting on the first term on the RHS of \arbitrary, we find:
\eqn\firstterm{
	\Qbar f_1(z, \bar z) \beta \ket{0}
	= i \bar \del f_1 \bThet\beta \ket{0}\ .
}
If $f_1(z)$ is holomorphic, this state is in the
kernel of $\Qbar$.  Furthermore, $\Qbar$ acts
on the third term on the RHS \arbitrary\ to give:
\eqn\thirdterm{
	\Qbar f_3(z) \wp\dag \ket{0} = i z f_3(z) \beta \ket{0}\ .
}
Finally, the second and fourth term on the RHS
of \arbitrary\ are Q-exact.

Thus, within cohomology,
\eqn\cohorelation{
	f_1(z) \simeq f_1 (z) + z f_3(z)\ .
}
The cohomology representatives live in
the ring $\IC[z]/\vev{z}$,
the local ring at $z=0$ (\cf\ \professorgriff).  There is a
single cohomology generator localized at $z=0$.  This is
our state, up to $\Qbar$ descendants.

We can see even more explicitly that
the string endpoint is stuck at the
origin and has a single Chan-Paton
factor by solving for the ground state
wavefunction of the worldsheet zero modes,
which is killed by both $\Qbar$ and $Q$.
To do this exactly we need the IR effective action for the
zero modes.  In this example the bulk
theory is a free scalar field theory
and the relevant perturbations lie entirely on the boundary.
If we assume that the RG flow does not add anything
to the bulk kinetic terms (as appears to be the case
in related work \refs{\boundaryflow}), and if the
boundary superpotential satisfies a nonrenormalization
theorem similar to that for the bulk superpotential, then the only
effect of the RG flow in the presence of
the relevant boundary perturbation \sheafysubcyclesuper\
will be to change the anomalous dimension of the
superpotential.  As the term \sheafysubcyclesuper\
is relevant along the RG flow, it will be dressed
with a factor $M^{\alpha}$, where $M\rightarrow\infty$
in the IR and $\alpha$ is positive.

Altering the supercharges accordingly, the ground state
should satisfy:
\eqn\qpsi{
\eqalign{
	0 &= \Qbar \ket{\psi_0} =
	\left( - i \bar \del f_1 - i M^\alpha z f_4 \right)
		\bThet \beta \ket{0}
	+ \left(- i \bar \del f_3 \right)
		\bThet \wp \dag \ket{0}
	+ M^\alpha \left( i z f_3 \right) \beta \ket{0}\cr
	0 &= Q \ket{\psi_0} = \left(- i f_1
		\bar z - i M^\alpha \del f_4 \right) \wp \dag\ket {0}
	+ M^\alpha \left(i \del f_2 \right) \beta \ket{0}
	+ \left( - i \bar z f_2 \right) \wp \dag \bThet \ket{0}\ .
}}
where we have ignored time derivatives since
$\Qbar=Q=0$ implies $H=0$.
The first equation in \qpsi\ requires that $f_3$ is holomorphic,
with support only at $z=0$; thus $f_3 = 0$.
Similarly, the second equation leads to $f_2 = 0$.
Rotational invariance implies that the lowest-energy
state $\ket{\psi_0}$ depends only on $(z \bar z)$.
Therefore,
\eqn\groundstate{
	\ket{\psi_0} \sim
	A(M) e^{ - M^\alpha |z|^2} \left( \beta \ket{0} +
	\wp \dag \Theta \dag \ket{0} \right)\ .
}
In the infrared, as $M\rightarrow\infty$,
the ground state will become a delta function
at $\CS = \{z=0\}$.  This has also been
observed in \kmmbsft.

\subsec{Unions and intersections}

In this section, we explain a number of ways to
combine two boundary linear sigma models to make another.
For argument, we discuss two transverse
hypersurfaces of a CY defined by
$S_1(\phi) = 0$ and $S_2(\phi) = 0 $ with
degrees $d_1$ and $d_2$.

\subsubsection{Intersections}

Here are {\it three} distinct ways to make the strings end on the
intersection $\left\{S_1 = S_2 = 0\right\}$:

\item{1.}
As described in \S2, add a boundary superpotential:
\eqn\intersectionpotential{
 W = \gamma_1 S_1 + \gamma_2 S_2 .
}
The vacuum energy $|S_1|^2 + |S_2|^2$ only vanishes on the 
intersection.  

\item{2.}  
The tensor product of two sheaves has support on the 
intersection of 
the support of the two sheaves.  Given the 
resolution of two sheaves by a sequence of line bundles, 
there is a formula for such a resolution of the tensor 
product whose derivation from the LSM is described in Appendix C.  
In this example, it gives:
\eqn\tensorexample{
\eqalign{
0 \to \CO & {\buildrel S_1 \over\longrightarrow} \CO(d_1) \to 0 \cr
&\bigotimes \cr
0 \to \CO & {\buildrel S_2 \over\longrightarrow}  \CO(d_2)\to 0 \cr
 &=\cr
0 \to \CO  {\buildrel \left
	 [ \matrix { S_1 \cr S_2 } \right ]  \over\longrightarrow}
	 \CO(d_1) & \oplus \CO(d_2) 
	{\buildrel \left[ -S_2, S_1 \right] \over\longrightarrow}
	\CO(d_1 + d_2) \to 0\ .}}
The field content is a neutral $\wp$, two $\beta$'s of charges $d_{1,2}$, and
a $\tilde \wp$ of charge $- (d_1 + d_2)$
with the constraints on
the $Q$'s and $Q\dag$'s determined by the sequence (\cf\ \hell).

\item{3.}  There is another sheafy description
of this intersection.
Take as boundary fields $\beta_{1,2}$ and $\wp_{1,2}$,
coupled via the boundary superpotential:
\eqn\sheafy{ W = \wp_1 \beta_1 S_1 + \wp_2 \beta_2 S_2 }
and project on separate boundary symmetries for each of
$\beta_1, \wp_1$ and $\beta_2, \wp_2$.
Away from the intersection the boundary system
cannot satisfy one or
the other charge projection while staying on the vacuum manifold.

\subsubsection{Unions}

Three analogous models for the union
$\left\{S_1=0\right\} \cup \left\{ S_2 =0\right\}$ are:

\item{1.} The scheme variables union is simply to use a
{\it single} boundary fermion of charge $ - d_1 - d_2 $
and add
\eqn\schemeyunion{ W = \gamma S_1(\phi) S_2(\phi) .}
The boundary energy is then $|S_1 S_2|^2$ and it vanishes
if one lies on {\it either} hypersurface.
This gives a scheme-theoretic description in the
following sense.
Notice that if $S_1$ and $S_2$ are the same, one gets
multiple Chan-Paton sectors.
In this sense, this boundary linear model
keeps track of the fact that a pair
of coincident branes is a non-reduced scheme.
Related observations were made in \erictom.

\item{2.} As with a single hypersurface, we can describe this 
in ``sheafy'' variables.  
By this we mean
replace $\gamma$ by a fermion $\beta$ and a boson $\wp$,
gauge the new boundary symmetry, and add
\eqn\sheafyunion{ W = \wp \beta S_1(\phi) S_2(\phi) .}

\item{3.} While tensoring sheaves intersects their supports, adding them
gives a sheaf whose support is the union.
Simply adding together
the sequences we find:
\eqn\directsumsequence{
\eqalign{
	0 \to \CO {\buildrel S_1 \over\longrightarrow} 
		&\CO(d_1) \to 0\cr
	&\bigoplus \cr
	0 \to \CO {\buildrel S_2 \over\longrightarrow} 
		&\CO(d_2) \to 0 \cr
	&=\cr
	0 \to \CO \oplus \CO {\buildrel [S_1, S_2] 
		\over\longrightarrow} 
	&\CO(d_1) \oplus \CO(d_2) \to 0 \ .
}}
The superpotential
is the same as \sheafy.
But if there is just a {\it single} charge projection, we obtain the
{\it union} of the two branes because
the string ends on one brane {\it or} the other.
So this is really the analog of the third way
of intersecting - we just do one charge projection instead of two.
This is also consistent with the fact that if only {\it one} of the boundary
symmetries is gauged, then there is one left over
which acts as the second $U(1)$ spacetime gauge symmetry
- the branes can move independently.

\newsec{Applications and consequences}

\subsec{Monodromy in closed-string moduli space}

The open string linear sigma model is an ideal tool
for thinking about transport of branes in closed
string moduli space.  Choose some boundary field
content to make a particular D-brane in the IR.
If we move
along a closed path in K\"ahler moduli space,
we come back to the same bulk LSM, but with
possibly different boundary data.
This determines
an action of the monodromy group on the
branes themselves, and not just on their charges.
Such a refinement of the action of the monodromy group
on branes has also been observed in the approach of
\refs{\dougcat,\categoryrefs,\asplaw} where branes are considered as
objects in the derived category.

In the case of the quintic, there
are three monodromy generators \refs{\candelas}.
At large volume, the imaginary
part of the complexified K\"ahler form --
represented in the linear model by the worldsheet
theta angle --
has periodicity $2\pi$.  A shift by $2\pi$
at large $r$ is a noncontractible loop
and should generate the ``large-volume monodromy.''
A loop about $r=0$ should generate
a monodromy which corresponds to the
monodromy about the conifold point in the
mirror CY.  Finally, at large negative $r$ a shift of
$\theta$ by $2\pi$ is also a noncontractible loop in the
closed string GLSM.  Since $r\to -\infty$
describes the Gepner point which is a $\IZ_5$ orbifold
point in the moduli space, this shift of $\theta$
should generate the $\IZ_5$ monodromy action about the Gepner point.

\bigskip
\noindent
{\it Large radius}

Consider a single D6-brane on the quintic, modeled
by a single neutral boundary fermion
which carries charge 1 under
the gauged boundary symmetry.
The pertinent terms in the boundary action are
\eqn\importantboundaryterms{
\CL_{bdy} \ni \left( { \theta \over 2 \pi} v_0 + j v_0 + (j_s - 1)a_0
\right)
}
where $v_0$ is the bulk worldsheet gauge field, $j$ is the bulk gauge
current, $j_s$ is the boundary symmetry current, and $a_0$ is the boundary
vector field in \bndrylagrange.

Now, shifting $\theta \to \theta + 2 \pi \alpha$ adds
to this
\eqn\thetashift{
\delta \CL = \alpha v_0,
}
(this is only gauge invariant if $\alpha$ is an integer 
multiple of $2 \pi$).
This is equivalent to shifting the bulk gauge
charges of all fields by their boundary charge:
\eqn\chargeshift{
 j \to j + \alpha j_s .
}

In our example,
this means that when $\alpha = 1$,
we can remove the
term \thetashift\ by giving the boundary fermion bulk gauge
charge $1$.
But this shifts the chern classes of the bundle; in particular, it adds
one unit of four-brane charge.  This is the expected large-radius
monodromy.

This is the right answer for any boundary field content.  Adding the
boundary symmetry generator to the gauge charge has the effect of
tensoring the bundle with the line bundle $\CO(1)$ over the
projective space.  This is precisely the effect of moving
the NS B-field through one period \refs{\candelas,\bdlr}.

\bigskip
\noindent
{\it Other monodromy generators}

An understanding of the other independent monodromy
generator requires an analysis of the effective
theory at the conifold singularity.
Work is in progress in this direction.  Here we restrict ourselves
to a few suggestive observations.

Firstly, even at large negative $r$,
$\theta \to \theta + 2\pi$ acts by adding the boundary current
to the gauge current.  So it seems that
the idempotence of the LG monodromy action
is related to the $\IZ_5$-valuedness
of the worldsheet gauge charge.

Secondly, it generates the expected monodromy on the branes
following the discussion in \refs{\diacdoug}.
A ``fractional brane'' in their model
corresponds to a brane whose Chan-Paton factors
have $\IZ_5$ charge under the orbifold group,
corresponding to the different irreducible representations
of $\IZ_5$.  The quantum symmetry at the orbifold point
rotates these irreps and so
shifts all of the charges by $1$ mod $5$.

Finally, the trivial representation
is believed to correspond to the D6-brane
at the Gepner point \refs{\diacdoug}.
This is the same description as at large
radius.  The fact that the D6-brane has
no monodromy about $r=0$ suggests that we
have the right description at the Gepner point.

\subsec{Degenerations and Singularities}

Singularities in the CFT moduli space are
especially important.  They provide
a window into nonperturbative physics,
and can give rise to the singularities of the spacetime superpotential
which are expected
in ${\cal N}=1$ supergravity \refs{\bagwit,\evaed}.
In the closed string GLSM, singularities of the CFT
appear at points in the GLSM moduli space where the
vacuum manifold becomes noncompact \phases.
This should also hold true in the open string case.
Furthermore, apparent singularities in the open
string moduli space, such as ``small instanton''
singularites, signal enhanced gauge symmetries
or the existence of a branch structure in the moduli space.

\bigskip
\noindent{\it Singularities in the closed string moduli space}

For closed strings,
singularities in the complex structure moduli space
occur when $G$ is not transverse;
then, as in \phases, there is a branch where $p$
and some of the $\phi$s diverge with zero energy cost.
If there are boundary terms of the form
\boundarysuper, where $S$ depends only on the bulk fields,
$\phi^i$ will not always be able to diverge consistent
with $S^A(\phi) = 0$, so the $D$-term keeps $p$ from diverging.
The boundary CFT will ${\it not}$
be singular (i.e. the D-brane physics
will be smooth) at many points where the closed string CFT would
be singular.

If $S = 0$ {\it is}\ consistent with the nonzero
$\phi^i$ necessary for the existence
of the non-compact $p$ branch,
then the full boundary CFT ${\it will}$ be singular.
In other words, there can be singularities in the relevant
boundary CFT as well as the closed string CFT
when the D-brane intersects the closed string singularity.

Singularities in the K\"ahler moduli space
occur when $\sigma$ has a noncompact branch.
For the closed GLSM on the quintic, this
occurs at $r=\theta=0$.  It is an interesting
question whether D-brane physics is singular
at this point -- see for example \matrixcy.
If the Chan-Paton factors for a given state
have nontrivial gauge charge then the answer
is uncertain.  When $\sigma$
is large all of the charged fields
have masses of order $|\sigma|^2$.
The charged boundary fermions create
a constant electric field in the bulk.
Competing effects exist:
on the one hand one expects screening of the bulk electric field
via
Schwinger pair production, on the other hand the mass
of the bulk charged fields (which must be pair-produced to
provide the screening) grows quickly down the
$\sigma$ branch.   It would be interesting to
disentangle the
physics of the potential singularity by performing
a delicate analysis of the IR limit (as was done
for closed strings in \evaed).

\bigskip
\noindent{\it Singularities from boundary fields?}

If the boundary field $\wp$ has no potential,
one might worry that a noncompact branch develops
and the CFT is singular.  For example,
in the Calabi-Yau phase, $\wp$ is usually frozen to zero by the term
$\sum_a \vert \wp\vert^2 \vert F^a\vert^2$, which gives $\wp$ a large
mass since the $F^a$ are generically nonvanishing at
$S=0$ in the Calabi-Yau.  However, one could
choose a bundle with singular points, where the
$F^a$ all vanish on $S=0$.  Even more uncomfortably,
in the Landau-Ginzburg phase $\phi=F=0$.  It would
be bizarre if the CFT was singular in an open set in the
moduli space.

However a $\wp$ branch does not exist.
The boundary charge projection gives a finite number of states
in the Hilbert space of boundary fields.
The $\wp$ field is a ``quantum dimension'' in the target space,
and cannot cause divergences in the path integral.
The integral over $a_0$ projects onto:
\eqn\specproj{j_s = \sum_a :\bbet_a \beta_a: -
	: \wp\dag \wp: = 1\ .}
Since the $\beta$s are fermionic, the positive contribution
in \specproj\ is bounded; therefore (when \specproj\
can be satisfied),
$\wp$ contributes
a finite volume factor to the path integral
and does not give any new branches to the path integral.

\bigskip
\noindent{\it Enhanced gauge symmetries}

Despite the absence of a $\wp$ branch,
there is significant physics when
a bundle degenerates.  If a locus $\CD$ exists such that
for $\phi^i_\star \in \CD$,
$F^a(\phi_\star) = G(\phi_\star) = 0$, then in addition
to the bundle $V$, we get a sheafy variables description of
a lower-dimensional brane~-- namely a sheaf with support over $\CD$.
The small instanton limit \smallinst\ of the D0-D4 system is
the classic example of such a degeneration.

In addition, we can sometimes tune parameters such that
additional global
symmetries arise, in addition to \specialsym\
which rotates all $\beta$'s and $\wp$'s
oppositely.  Then additional spacetime gauge
symmetries should appear.

\bigskip
\noindent
{\it Small Instanton ``Singularities''}

The ``sheafy variables'' construction of
D-branes allows us to easily study transitions
between branches of D-brane moduli spaces.
First we will study two D2-branes filling orthogonal
complex lines in a flat $\IC^3$.  The ``Higgs branch''
can occur when they lie in a common $\IC^2$
and can be deformed into a single D2-brane;
the ``Coulomb branch'' occurs when they are separated
along the complex line orthogonal to both of them.
If we replaced $\IC^3$ by $T^6$ this would
be the T-dual of the small instanton singularity.
We will then proceed to a direct study of small instantons in
the D0-D4 system.

Consider the linear model which flows to the Coulomb branch.
We take as boundary fields $\b,\bt, \wp_{1,2}, \b\pri,\bt\pri$.
We wish to describe a background with one two-brane located at
$z_1 = 0, z_3 = 0$, and the other at $z_2 = 0, z_3 = a$.

Using the ``sheafy variables''
description, the complex defining this configuration is
$$
\eqalign {
\left (
\matrix{
        z_1            \cr
0   \to  \CO \to  \CO  \to  0 \cr
   \otimes   \cr
     z_3    \cr
0   \to  \CO \to  \CO  \to  0 \cr
}
\right )
&\bigoplus
\left (
\matrix{
        z_2            \cr
0   \to  \CO \to  \CO  \to  0 \cr
   \otimes   \cr
     z_3-a    \cr
0   \to  \CO \to  \CO  \to  0 \cr
}
\right )
\cr
&=
}
$$
$$
\left (
\matrix{
&  & & \left [ \matrix { z_1\cr z_3 } \right ]   & &
 \left [ \matrix z_3 , -z_1  \right ]  & & & \cr
 0  &\to &\CO &\to  &\CO\uu{\oplus 2}  &\to  &\CO  &\to  &0
}
\right )
$$
$$
\bigoplus
$$
$$
\left (
\matrix{
 &  & & \left [ \matrix { z_2\cr z_3-a } \right ]   & &
 \left [ \matrix z_3-a , -z_2  \right ]  & & & \cr
0  &\to &\CO &\to  &\CO\uu{\oplus 2}  &\to  &\CO  &\to  &0
}
\right )
$$
$$
= \matrix {
& & & \left [ \matrix { z\ll 1 & 0 \cr z\ll 3 & 0 \cr 0 & z\ll 2\cr
0 & z\ll 3 - a } \right ] & & \left [ \matrix
{ z\ll 3 & -z\ll 1 & 0 & 0 \cr 0 & 0 & z\ll 3 -a & -z\ll 2}
\right ]  & &
\cr
0 & \to &\CO\uu{\oplus 2} &\longrightarrow&\CO\uu{\oplus 4}&
\longrightarrow&\CO\uu{\oplus 2}&\to&0
}
$$
In other words, the deformed chiral constraints are
\eqn\deformedconstraintsforhc{
\matrix{
Q\dag \b = 0 & Q\dag \wp\dag\ll 1 = z\ll 1 \b & Q\dag \wp\dag\ll 2
= z\ll 3 \b & Q\dag \b\pri = z\ll 3 \wp\dag\ll 1 - z\ll 1 \wp\dag\ll 2
\cr
Q\dag \bt = 0 & Q\dag \wpt\dag\ll 1 = z\ll 2 \bt & Q\dag \wpt\dag\ll 2
= (z\ll 3-a) \bt & Q\dag \bt\pri = (z\ll 3 -a)\wpt\dag\ll 1 - z\ll 2 \wpt\dag
\ll 2
}
}
and the on-shell supersymmetry transformations are
\eqn\onshelltransfsforhc{
\matrix {
Q\dag \b\dag =  -z\ll 1 \wp\ll 1 - z\ll 3 \wp\ll 2
& Q\dag \wp\ll 1 =  -z\ll 3 {\b\pri}\dag & Q\dag \wp\ll 2
=  z\ll 1 {\b\pri}\dag  & Q\dag {\b\pri}\dag = 0
\cr
Q\dag \bt\dag = - z\ll 2 \wpt\ll 1 - (z\ll 3 -a)\wpt\ll 2
& Q\dag \wpt\ll 1 = -(z\ll 3 -a)\bt^{\prime\dagger} & Q\dag \wpt\ll 2
= z\ll 2 \bt^{\prime\dagger}   & Q\dag \bt^{\prime\dagger} = 0
}
}

One can search for marginal deformations of the system in one of two
ways.  Either one can look for candidate superpotential terms annihilated
by $Q\dag$ (using the unperturbed EOM and values for the auxiliary fields);
or one can look for consistent perturbations of the off-shell
deformed chiral constraints and add no explicit superpotential term.
The two approaches yield equivalent results; here we will take the latter.

We are interested in deformations which preserve the orientation of the branes
and more generally the structure of the configuration at infinity,
so we only
allow deformations of the maps of the complex which are constant
independent of $z$.  Deformations linear in $z$ would generally alter the
orientation of the branes, and higher-order polynomials would cause even
more drastic changes in the brane geometry.

For $a\neq 0$ there are no constant deformations of the system other
than the obvious ones corresponding to moving the twobrane moduli
around on the Coulomb branch.

When $a$ vanishes, however, one can find other consistent perturbations of the
deformed chiral constraints, or equivalently deformations of the
complex preserving nilpotence of the differential $Q\dag$.

Specifically, for vanishing $a$ the most general set of constants one can
add to the maps of the complex is:
\eqn\mostgeneraldefs{
\matrix {
& & & \left [ \matrix { z\ll 1 & \e \cr z\ll 3 & 0 \cr \e\pri & z\ll 2\cr
0 & z\ll 3 } \right ] & & \left [ \matrix
{ z\ll 3 & -z\ll 1 & 0 & -\e \cr 0 & -\e\pri & z\ll 3 & -z\ll 2}
\right ]  & &
\cr
0 & \to &\CO\uu{\oplus 2} &\longrightarrow&\CO\uu{\oplus 4}&
\longrightarrow&\CO\uu{\oplus 2}&\to&0
}
}

We will now show that for $a=0$ the effect
of this perturbation will be to merge the two branes into a single brane
covering the locus $z_3=0, z_1z_2 = \e\e\pri$.  After adding this deformation 
of the complex, the deformed chiral constraints become 
\eqn\deformeddeformedchiralconstraints{
\matrix{
Q\dag \b = 0 & Q\dag \wp\dag\ll 1 = z\ll 1 \b + \e\ll 1\bt & Q\dag \wp\dag\ll 2
= z\ll 3 \b & Q\dag \b\pri = z\ll 3 \wp\dag\ll 1 - z\ll 1 \wp\dag\ll 2
-\e\wpt\dag\ll 2
\cr
Q\dag \bt = 0 & Q\dag \wpt\dag\ll 1 = z\ll 2 \bt + \e\pri \b
& Q\dag \wpt\dag\ll 2
= z\ll 3 \bt & Q\dag \bt\pri =  z\ll 3\wpt\dag\ll 1 - z\ll 2 \wpt\dag
\ll 2 -\e\pri\wp\dag\ll 2
}
}
and the on-shell supersymmetry transformations are
\eqn\deformedonshellsusy{
\matrix {
Q\dag \b\dag =  -z\ll 1 \wp\ll 1 - z\ll 3 \wp\ll 2 - \e\pri\wpt\ll 1
& Q\dag \wp\ll 1 =  -z\ll 3 {\b\pri}\dag & Q\dag \wp\ll 2
=  z\ll 1 {\b\pri}\dag  + \e\pri \bt\uu{\prime\dagger}
 & Q\dag \b\uu{\prime\dagger} = 0
\cr
Q\dag \bt\dag =  - z\ll 2 \wpt\ll 1 - z\ll 3 \wpt\ll 2 - \e\wp\ll 1
& Q\dag \wpt\ll 1 = - z\ll 3 \bt^{\prime\dagger} & Q\dag \wpt\ll 2
= z\ll 2 \bt^{\prime\dagger}+\e \b\uu{\prime\dagger}
  & Q\dag \bt^{\prime\dagger} = 0
}
}

The condition for the $Q$ and $Q\dag$ variations of all $\wp$ fields
to vanish is
\eqn\vacummeqnsforcoulombbranch{
\eqalign{
z\ll 3 (\b,\bt,\b\uu{\prime\dagger},\bt\uu{\prime\dagger}) &= 0 \cr
 (z\ll 1 z\ll 2 - \e\e\pri)(\b,\bt,\b\uu{\prime\dagger},
\bt\uu{\prime\dagger}) &=0.
}}
So we see that this deformation moves the support of the brane
to the irreducible variety defined by
\eqn\irreducible{ z_3 = 0, ~~ z_1 z_2 = \epsilon \epsilon \pri }
which represents the Higgs branch of this D2-D2 system.

There is a shortcut to finding the locus of the twobrane.  If one takes
a generic point in the base space, the matrices defining the complex
are all of full rank.  Their rank is reduced, leading to nonzero
cohomology, exactly when all two by two subdeterminants vanish.  This
occurs at the locus $z_3 = 0, z_1 z_2 = \epsilon\epsilon\pri$.
\bigskip
\noindent
{\it Higgs-Coulomb transition in the D0-D4 system}

The merging of two $D2-$branes is $T-$dual to the Coulomb-Higgs
transition in the 0-4 system.  We can also describe this transition directly.

A linear sigma model for a zerobrane separated from a pair of fourbranes is
\eqn\lsmfordzerodfour{
\matrix{
 &   &   & d_1 &              & d_2&             &d_3&   &     & \cr
0&\to&\CO& \to &\CO^{\oplus 5}&\to&\CO^{\oplus 5}&\to&\CO& \to &0
}
}
with
\eqn\done{
d_1 \equiv \left [ \matrix { z_1 \cr z_2 \cr z_3 - L \cr 0 \cr 0} \right ]
}
\eqn\dtwo{
d_2 \equiv \left [ \matrix { 0 & z_3 - L & - z_2 & 0 & 0\cr
                             L - z_3 & 0 & z_1 & 0 & 0 \cr
                             z_2 & - z_1 & 0 & 0 & 0 \cr
                             0 & 0 & 0 & z_3 & 0 \cr
                             0 & 0 & 0 & 0 & z_3 } \right ]
}
\eqn\dthree{
d_3 \equiv \left [ \matrix {   z_1 , & z_2, & z_3-L &0, &0 } \right ].
}

For generic $L$ the only marginal operators one can add simply correspond
to shifts in the positions of the branes.  However for $L = 0$ one can
deform the complex to:
\eqn\deformeddone{
d_1 \equiv \left [ \matrix { z_1 \cr z_2 \cr z_3  \cr -B_1 \cr -B_2} \right ]
}
\eqn\deformeddtwo{
d_2 \equiv \left [ \matrix { 0 & z_3  & - z_2 & 0 & 0\cr
                              - z_3 & 0 & z_1 & 0 & 0 \cr
                             z_2 & - z_1 & 0 & C_1 & C_2 \cr
                             0 & 0 & B_1 & z_3+ E_{11} & E_{12} \cr
                             0 & 0 & B_2 & E_{21} & z_3 +E_{22}} \right ]
}
\eqn\deformeddthree{
d_3 \equiv \left [ \matrix {   z_1 , & z_2, & z_3, &-C_1, &-C_2 } \right ]
}
for $B,C,E$ which satisfy
\eqn\constraintsondef{
\left [ \matrix {  C_1,& C_2 }\right ]
\left [ \matrix {  B_1\cr B_2 }\right ]
= \left [ \matrix {  E_{11} & E_{12} \cr E_{21} & E_{22} }\right ]
\left [ \matrix {  B_1\cr B_2 }\right ]
=\left [ \matrix {  C_1,&C_2 }\right ]
\left [ \matrix {  E_{11} & E_{12} \cr E_{21} & E_{22} }\right ] =0
}
The general solution to these constraints consists of a unit
doublet $(u_1, u_2)$ and three complex numbers $b,c,e$:
\eqn\solutiontoconstraints{
\left [ \matrix {  B_1\cr B_2 }\right ] =
b \left [ \matrix {  u_1\cr u_2 }\right ]; ~~
\left [ \matrix {  C_1,& C_2 }\right ] =
c \left [ \matrix {  u_2,& -u_1 }\right ]; ~~
\left [ \matrix {  E_{11} & E_{12} \cr E_{21} & E_{22} }\right ] =
e \left [ \matrix {  u_1 u_2 & -u_1^2 \cr   u_2^2 & - u_1 u_2 }\right ]
}
We will now assume generic values of $b,c,e,u$.  
There is never cohomology at the first or last node; neither $d_1$
nor $d_3$ ever has a kernel.  $d_2$ always has determinant zero, which
is to be expected since $d_1$ has one-dimensional image, so $d_2$ must
have at least one-dimensional kernel.

There will be cohomology at the second and third nodes
only when the dimension of the kernel of $d_2$
is two or higher, the criterion for which is the vanishing of
all twenty-five 4 by 4 subdeterminants of $d_2$.
We find that the matrix of subdeterminants is
\eqn\cofactormatrix{
z_3^2\cdot
\left [ \matrix { -bcu_1 u_2, & bcu_1^2, & c u_1 z_3, & - cu_1z_2, & c u_1 z_1
            \cr -bc u_2^2, & bc u_1 u_2 ,&c u_2 z_3, & -c u_2 z_2 & c u_2z_1
            \cr -b u_2 z_3 & b u_1 z_3 & z_3^2 & -z_2 z_3 & z_1 z_3
            \cr b u_2 z_2 & -b u_1 z_2 & - z_2 z_3 & z_2^2 & - z_1 z_2
            \cr -b u_2 z_1 & b u_1 z_1 & z_1 z_3 & - z_1 z_2 & z_1^2 }
\right ]
}
Specifically, it factorizes as $z_3^2$ times a matrix which never vanishes.
(The entry in the upper left hand corner, for instance, is constant
and nonzero).
This
means that there are two fibers along the locus $z_3 = 0$ (rather than a single
fiber, since the zero
is doubled), and no
extra fibers at special subloci of $z_3 = 0$.
The interpretation is that we have dissolved a zerobrane into a smooth
instanton field in the pair of fourbranes.
(Note that the value of $e$ drops out of the subdeterminants
and we can set it to zero in what follows.)

We can see this even more directly by examining the cohomology restricted
to $z_3=0$.  Setting $e=0$ and computing the kernel of $d_2$ we have
\eqn\almostadhm{
d_2 \equiv \left [ \matrix { 0 & 0 & - z_2 & 0 & 0\cr
                              0 & 0 & z_1 & 0 & 0 \cr
                             z_2 & - z_1 & 0 & C_1 & C_2 \cr
                             0 & 0 & B_1 & 0 & 0 \cr
                             0 & 0 & B_2 & 0 & 0} \right ]
\left [ \matrix { \b_1 \cr \b_2 \cr \b_3 \cr \b_4 \cr \b_5 } \right ] =0
}
\eqn\partofadhm{
\Leftrightarrow
\left [ \matrix { z_2, & - z_1, & C_1, & C_2 } \right ]
\left [ \matrix { \b_1 \cr \b_2 \cr \b_4 \cr \b_5 }
\right ] = \b_3 = 0
}
Furthermore, the massless $\b$'s are in the kernel of $d_1^\dagger$,
restricted to $z_3 = 0$.  That is,
\eqn\otherpartofadhm{
\left [ \matrix { z_1^*, & z_2^*,  & B_1^*, & B_2^* } \right ]
\left [ \matrix { \b_1 \cr \b_2 \cr \b_4 \cr \b_5 } \right ] =0
}

In other words, the vector $[\b_1, \b_2, \b_4, \b_5]$ is a harmonic
representative of the cohomology of the complex
\eqn\adhmcomplex{
\matrix{
 &   &   &  \left [ \matrix { z_1 \cr  z_2 \cr  B_1 \cr B_2 } \right ]
 &
 & \left [ \matrix { z_2, & - z_1, & C_1, & C_2 } \right ]   &  & &
\cr
0&\to&\CO& \to &\CO^{\oplus 4}& \to & \CO& \to &0
}
}
with
\eqn\adhmeqn{
\left [ \matrix {  C_1,& C_2 }\right ]
\left [ \matrix {  B_1\cr B_2 }\right ] = 0
}

But
the ADHM construction of a single $U(2)$ instanton in $\IR^4$
works precisely by defining a holomorphic bundle as the cohomology of
this same complex.
So we have directly demonstrated the ability
of a D0 brane to dissolve into D4 branes by open string worldsheet
arguments, without any reference to the brane worldvolume gauge theory.

This whole computation should generalize without undue complication to
the process of dissolving $k$ instantons in $N$ fourbranes to make a smooth
$k$-instanton field in a $U(N)$ gauge theory on $\IR^4$.

\subsec{Marginal stability transitions: A local worldsheet model}

While many of the quantities which are reliably computed in the
open string LSM are independent of the worldsheet FI parameter,
it is clear that important aspects of the physics of B-type
D-branes do depend on
the K\"ahler moduli.  The most striking example is marginal stability
transitions, which occur at special loci in the K\"ahler moduli space.
Determining when a transition occurs for a given set of brane charges
is a delicate problem which has come under intense recent investigation
(see e.g. \refs{\bdlr,\mrdstab,\denef,\dougcat}).
Since the topological B-model is insensitive to the K\"ahler parameters,
the occurence of such transitions is not transparent in the linear model.
However, we ${\it can}$ understand the local physics of these transitions in the
linear model in a simple way.
The local model we present below is a good candidate to
describe the generic worldsheet behavior in the vicinity of such a transition.

A local model for the ${\it spacetime}$ physics of a
D-brane undergoing a marginal stability
transition is the Fayet model \Fayet.
This is easily seen in the A-model
\refs{\shamitjohn,\joyce}, where the physics can be reduced to
a problem involving intersecting branes.\foot{In
the B-model, these transitions are related to a chamber
structure in the moduli space of stable bundles
\sharpe.}
The Fayet model is a four-dimensional $\CN=1$ field theory
with a $U(1)$ gauge symmetry and a single charged field.
(In the brane system, there is also a center of mass vector
field gauging a $U(1)$ under which all fields are neutral.)  The
D-term takes the form
$$ D = |\phi|^2 - \xi .$$
When the FI parameter $\xi$ is positive, there is a
supersymmetric vacuum at nonzero $\phi$; the gauge
symmetry is broken by this vev and there is a mass gap.
For $\xi < 0$, there is no way to make $D$ vanish; supersymmetry
is broken, and the gauge symmetry is preserved.  Classically,
the gauge multiplet $( A_\mu, \lambda_\alpha )$, and
the fermion partner $\psi_\alpha$ of $\phi$ are all
massless, while $\phi$ is massive.
In realizations of this transition in B-type brane systems,
$\xi$ is to be identified with a nontrivial function of the
K\"ahler parameter $t = {\theta\over 2\pi} + ir$.  There is a supersymmetric brane for
$\xi > 0$ and the brane decays on the locus $\xi(t,t \dag) = 0$
in K\"ahler moduli space.

This simplest model of the spacetime physics is
reproduced by the simplest possibility in the LSM.
Add a brane-antibrane pair to the linear model -
\ie\ add a boundary chiral multiplet $\wp$ and a
boundary fermi multiplet $\beta$.

To model the behavior when $\xi$ is small and positive,
add the operator
\eqn\obvioustachyon{f(\xi) \int d\theta \wp \beta}
to the boundary
action ($f$ is some nontrivial function of $\xi$, which vanishes
when $\xi = 0$).
The $Q\dag$ complex then takes the form
\eqn\wouldbeQcomplex{
0 \to \left\{ \beta \dag \wp \dag \right\} \to
\left\{ \matrix{   \wp \dag \wp \cr
		   \beta\dag \beta }
\right\}
\to \left\{ \beta \wp \right\} \to 0.
}
At generic points when $\xi > 0$, $f$ is non-vanishing.
The last map from R-charge $0$ to R-charge $1$ is then onto,
and there is no cohomology beyond the center of mass vector
at R-charge $0$.
We have a supersymmetric vacuum with the mass gap (for modes
other than the decoupled $U(1)$) that we expect.
The R-symmetry of the CFT to which we flow
must preserve \obvioustachyon, so the
operator $\beta \wp$ has unit R-charge.

Now consider the behavior at $\xi = 0$.  Since $f(\xi) = 0$ at this point,
the maps in \wouldbeQcomplex\ degenerate
and we get a cohomology generator at R-charge zero
(in addition to the center-of-mass gaugino
$\bf{1} = \beta \dag \beta - \wp \dag \wp $) which
we identify as a new gaugino.  Its image under spectral flow
is a new massless vector.  There is also new cohomology
at R-charge $1$ of the form $\beta \wp$ which we identify with
the vertex operator for $\psi_\alpha$, the fermi
component of the charged chiral multiplet.
The R-charge of this operator is still unity, and so
spectral flow generates from it the vertex operator for
massless scalar $\phi$.
One can show, using the equation
\eqn\chargecoupling{
  [ \beta \dag \beta , \wp \beta ] = - \wp \beta,
}
that the string created by this vertex operator indeed carries unit
charge under the new vector field, as expected.

Now, what happens as one moves past the spacetime transition point, to the
$\xi < 0$ region of parameter space of the Fayet model?
At $\xi = 0$, the model enjoys an extra unbroken $U(1)$ global symmetry -- one
can rotate $\beta$ and $\wp$ independently.
There is no longer a unique candidate for the $U(1)_R$ symmetry which
appears in the IR ${\cal N}=2$ superconformal algebra.
We can hypothesize that when $\xi$ is made slightly negative,
the R-charge is a linear combination of the two $U(1)$s under which
$\beta \wp$ has charge $>1$.
We then find that the scalar $\phi$ obtains
a tree-level mass since the conformal weight of the image
of $V_\psi$ under spectral flow will be different from $1$.
The operator $\beta \wp$ is still in the $Q\dag$
cohomology, so $\psi_\alpha$ remains massless.

This scenario will be realized in the following situation.
When $\xi > 0$, the $\phi$ field has two gauge-inequivalent vacua (the true vacuum,
and the tachyonic vacuum at $\langle \phi \rangle = 0$).
Associated with these two vacua are two different boundary
CFTs.  In the CFT of the supersymmetric vacuum, where
the tachyonic perturbation, $\int d\theta \beta \wp$, has been turned on,
the conserved R-charge must be such that $\beta \wp $ has unit R-charge.
\smallfig\rcharges{
The R-charge, $q_R$, of the operator $\wp \beta$ varies
with the spacetime FI coefficient $\xi$.  The dashed
line indicates the conserved R-charge in the CFT of the
tachyonic vacuum at $\vev{\phi} = 0$.
}
{\epsfxsize1.7in
\epsfbox{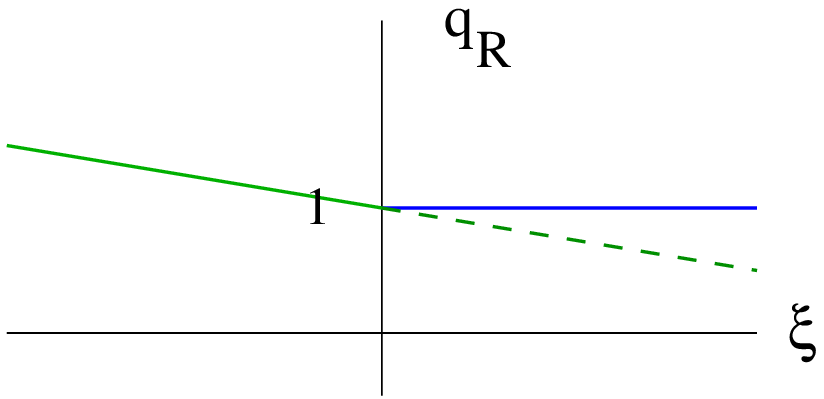}}
In the unperturbed CFT of the tachyonic $\langle \phi \rangle = 0$ vacuum,
there is an extra global boundary $U(1)$ symmetry
under which (if necessary by adding to it a multiple of the
gauged boundary current) only $\beta$ is charged.  Therefore the
conserved R-charge in the IR theory can be some linear combination
of the bulk R-current and the $\beta$ number current with $r$-dependent
(and hence $\xi$-dependent) coefficients determined by the boundary
RG flow.  It is this gauge-symmetric vacuum (and hence this CFT)
which describes the brane at $\xi < 0$, and so the R-charge of
$\wp \beta$ can vary with $\xi$ when $\xi$ becomes negative.

It would be very interesting to understand in detail, from a microscopic
point of view, the appearance of this local model in various
D-brane decay processes.
It is quite plausible that not only generic marginal stability transitions,
but also generic variations of the worldvolume spectrum on a given D-brane,
can be accomplished through the judicious addition of brane/anti-brane
pairs in this manner.

\newsec{Massless worldvolume fields}

\subsec{Generalities}

In this section we 
compute the massless worldvolume
spectrum in a particular example.  Our linear model
flows to the worldsheet-supersymmetric Ramond sector of strings ending
on these branes, and it is the
states in this sector which are annihilated by $L_0$ that we determine.
As in \refs{\dougcat,\asplaw,\diac} it should be understood that we are
looking at branes in the topological B-model, and the spectrum of
strings we are computing corresponds to the massless open (fermionic)
strings stretching between such branes.  The relation of these
``topological'' branes to physical branes is not always straightforward:
the supersymmetric branes in the topological model satisfy the
physical F-flatness conditions but not necessarily the physical
D-flatness conditions.  The question of which topological branes
satisfy the latter condition is equivalent to understanding marginal
stability, and loci of marginal stability are not manifest in the linear
model (and indeed still need to be determined on a more or less case
by case basis).
In some special cases, e.g. half-supersymmetric D-branes on $K3$, the
enhanced supersymmetry makes it easier to infer properties of the
physical branes from the topological model, so after
some generalities we will specialize to an
example involving branes on $K3$.

To find the massless open string states, we can study
the supersymmetric ground states in the Ramond sector.
In this sector, the two unbroken supercharges
$Q$ and $Q \dag$ anti-commute to the Ramond-sector
Hamiltonian:
\eqn\comm{ \{ Q,Q \dag \}~=~2L_0 }
By \comm\ and standard results
in Hodge theory, we can find the supersymmetric ground states by computing the
$Q \dag$ cohomology.
Since the FI terms are $Q$ exact, we can ignore their
effects in this computation.

We will find it convenient to
compute the cohomology of $Q \dag$ acting on {\it operators}.
\smallfig\ops{In the conformal limit, emission of a string stretched between
branes is the same as insertion of a boundary-condition-changing vertex operator,
$V_{ab} \sim \beta_b \dag \beta_a$.}{\epsfxsize3.0in\epsfbox{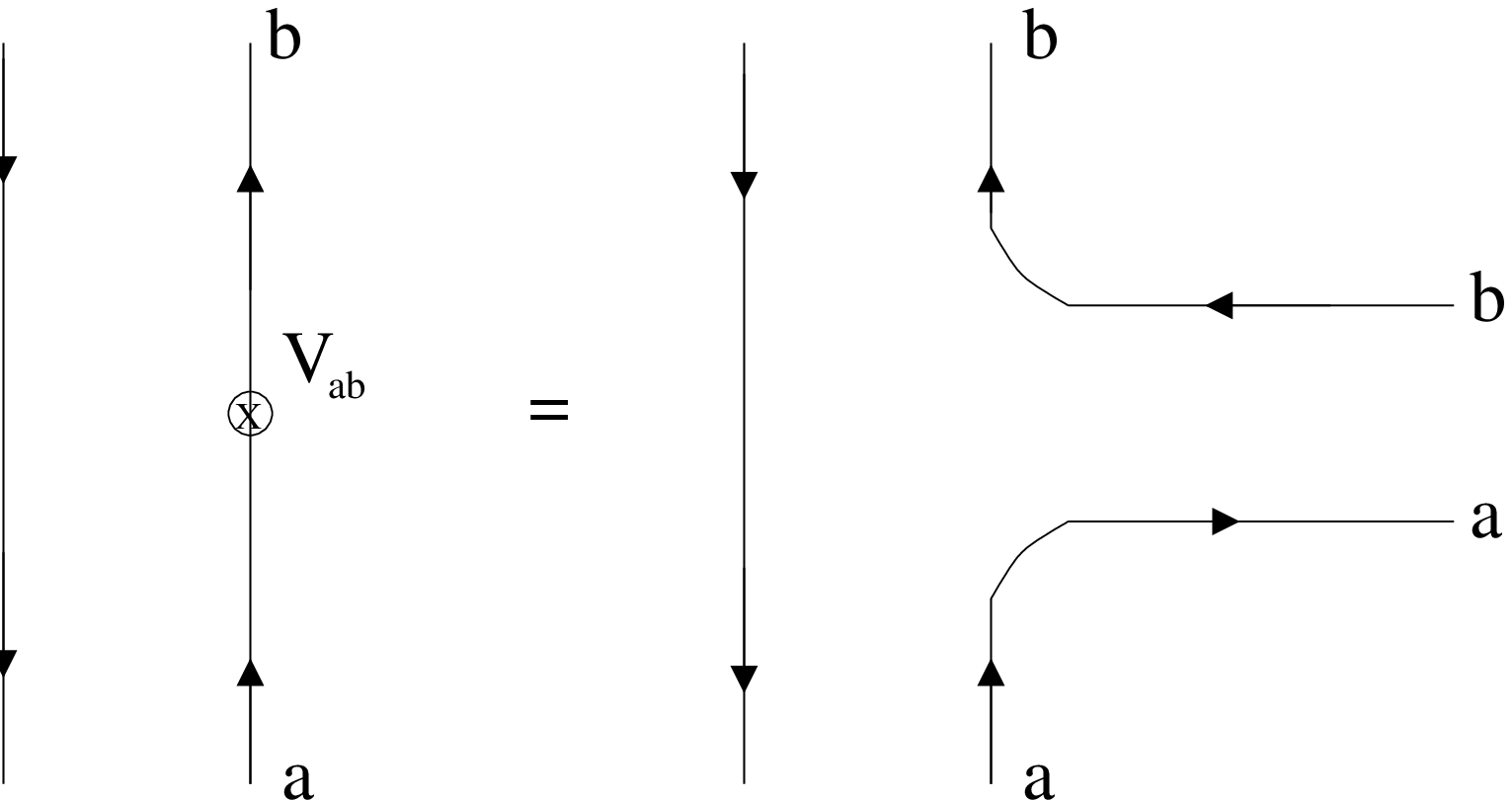}}
\noindent
Since, as illustrated in \ops, there is a 1-1 correspondence between
the open string states and associated boundary condition changing operators,
this does not constitute a loss of generality.

\subsec{Example: A bundle on $K3$}

\hbox{\vtop{\hsize=2.625in
\def\tablerule{\omit&
\multispan{8}{\tabskip=0pt\hrulefill}&\cr}
\def\tablepad{\omit&
height3pt&&&&&&&&\cr}
$$\vbox{\offinterlineskip\tabskip=0pt\halign{
\hskip-.25in
\strut$#$\ \ &
\vrule#&\ \ \hfil $#$ \hfil\ \ &\vrule #&\ \
\hfil $#$ \hfil\ \ &\vrule #&\ \
\hfil $#$ \hfil\ \ &\vrule #&\ \
\hfil $#$ \hfil\ \ &\vrule #\cr
&\omit&\hbox{Field}&\omit&q_G&\omit&q_R&\omit&q_S&\omit\cr
\tablerule\tablepad
&&\phi^i&& 1 && \half && 0 &\cr
\tablepad\tablerule\tablepad
&&\Theta^i, \eta^i && 1 && -\half && 0 &\cr
\tablepad\tablerule\tablepad
&&p&&-4&&0&& 0& \cr
\tablepad\tablerule\tablepad
&&\Theta_p, \eta_p&&-4&&-1&& 0& \cr
\tablepad\tablerule\tablepad
&&\beta_a&&1&& -\half && 1&\cr
\tablepad\tablerule\tablepad
&&\wp&&-4&&0 && -1&\cr
\tablepad\tablerule
\noalign{\bigskip}
\noalign{\narrower\noindent{\bf Table 1:}
The charges of the relevant fields.}
 }}$$}
\vtop{\advance\hsize by -2.625in
Our example consists of three D4-branes on the quartic K3,
with the bundle
$V$ defined by the sequence \seq\ with $m = 4$ and $\{n_a \} = \{1,1,1,1\}$.
There is a boundary superpotential coupling
$\int d \theta \beta_a F^a(\phi) \wp$.
The table at left displays the gauge charges, R-charge, and
boundary global charges of all of the relevant fields.
The bulk vector is neutral under all of
these transformations.
}}

In the GLSM for this simple monad on K3,
our supercharge takes the form
\eqn\GLSMsuperc{
\eqalign{
&~~~~~~Q \dag = Q_{\rm bulk}\dag  + Q_{\partial} \dag  \cr
& = \int dx ~
\left\{
\nabla_0 \phi_i \Theta^{i \dagger} + \nabla_1 \phi_i \eta^{i \dagger}
                    + 2 p \del_i G \eta^i +
	\nabla_0 p \Theta_p \dag + \nabla_1 p \eta_p\dag + 2 G \eta_p
\right.
\cr
&+ \lambda_+
\left(- \del_+ \sigma + {1 \over \sqrt{2}} \left(i D + {v_{+-} \over \sqrt 2}\right) \right)
	+ \lambda_-
\left( \del_- \sigma \dag -
{1 \over \sqrt{2}} \left( i D - {v_{+-} \over \sqrt 2}\right) \right)
\cr
&\left.
~~~~~~~~+ \delta(\dS) \left( \wp \beta_a F^a(\phi) + \eta_p\dag p  \right)
\right\}.
}}

\item{$\bullet$}
We will only be interested
in operators which are invariant under the gauged special boundary
symmetry (so that they map states in the charge-one sector back into
the same sector).  Furthermore, if we are not going to
restrict ourselves to a phase of the theory
where the bulk $U(1)$ is higgsed
and integrate out the bulk gauge field, we must throw away operators which
are not $U(1)$ invariant.  We want to find the spectrum of
massless
spacetime fields.  The fermionic parts of such multiplets
arise from Ramond vertex operators whose internal
parts have R-charge $0$ or $1$.  The former map via
spectral flow to NS-sector operators of conformal weight $0$
and hence lead to spacetime vectors.  Ramond states of unit
R-charge flow spectrally to NS operators of unit dimension
which give the scalar components of spacetime chiral multiplets.

\item{$\bullet$}
The equation of motion of the boundary vector field $a_0$ is imposed as
an operator equation of motion.  So we impose the boundary Gauss' Law,
\eqn\gauss{   :\beta_a \dag \beta_a: - : \wp \dag \wp: = \pm 1 }
where the $\pm$ depends on the orientation of the boundary component,
on our operators.

\item{$\bullet$}
In order to be boundary-symmetry-invariant, an operator has to contain an
even number of boundary fields.  Operators without any boundary fields
are just
bulk operators restricted to the boundary and we ignore them.
Included among these is the identity operator, which
creates the center-of-mass vector.
We claim that operators containing four or more
boundary fields do not create any states beyond
those made by operators bilinear in the fiber fields.
We can see this by making the state-operator correspondence more
explicit as follows.

The only subtlety involved is
the vacuum degeneracy (present even in the NS sector)
arising from the boundary fermion and boson zero modes.
To resolve this, we choose a reference state,
\eqn\refstate{ \ket{0}_+ \otimes \ket{0}_- .}
Here $\ket{0}_\pm$ are vacua of the two ends of the string
satisfying
\eqn\vacuumdefinition{ 0 = \beta_{a} \ket{0}_- =
\wp \dag \ket{0}_- = \beta_{a} \dag \ket{0}_+ =
\wp \ket{0}_+ .}
So we have arbitarily picked out a state
which is not in the boundary
Hilbert space 
of the ends of the string because it
does not satisfy
the boundary charge projection.
The state corresponding to an operator is
obtained by acting with that operator on
this vacuum and projecting
onto the subspace satisfying the charge projection.
For example, an operator of the
form $\beta_a \dag \beta_b $ makes the state with
the $-$ end of the string in sector $a$,
and the $+$ end of the string in sector $b$.
Acting on this state, the only independent
boundary operators invariant under the boundary symmetry
are of the form
$$ \wp \dag \beta \dag, \wp \dag \wp, \beta \dag \beta, \wp \beta.$$
Operators which are least quartic in the fiber fields
are made by acting with one of these operators again.
Using the charge projection \gauss\
this will always give zero or a state created by
a quadratic operator.

\item{$\bullet$}
It is convenient to divide up $Q \dag$ into
\eqn\zigzag{
 Q \dag = Q\dag_0 + Q \dag _1 
}
where $Q\dag_0$ includes only the parts depending
on derivatives of bulk fields.
By a zig-zag argument of the type appearing in \kw\
we can compute the cohomology of $Q \dag$ by computing the
cohomology of $Q\dag_1$ in the cohomology of $Q\dag_0$.
This tells us that we can
leave out any non-holomorphic dependence on $\phi$
and we
can leave out $\Theta$'s and their daggers because
any operator with $\Theta$- or $\phi \dag$-dependence is
a $Q \dag$ descendant.

\item{$\bullet$}
We discard the vector multiplet from our cohomology
calculation.  It is massive at large $|r|$, where its
effect is to impose Gauss' law and its supersymmetric
completions.  Combined with the fact that
the result of this calculation is independent of
worldsheet FI terms (which are $Q\dag$ descendants), this
means that we may safely neglect it.
A more rigorous justification for this awaits future work.

\item{$\bullet$}
Operators containing the bulk field $p$, but not containing
its partner $\eta_p\dag $ are in the image of the
supercharge because of the term
\eqn\killingp{
 \delta(\dS) \eta_p \dag p
}
added to the supercharge to solve the Warner problem.

\bigskip

Schematically, the structure of the action of $Q\dag$ on the
pertinent operators is:
\eqn\Qcomplex{
0 \to \left\{ R^a_{(l)}(\phi) \beta_a \dag \wp \dag \right\} \to
\left\{ \matrix{  S_{(l+3)}(\phi) \wp \dag \wp \cr
		  S^{ab}_{(l+3)}(\phi) \beta_a \dag \beta_b }
\right\}
\to \left\{ T^a_{(l+6)}(\phi)\beta_a \wp \right\} \to 0,
}
where the notation means that $R^a_{(l)}(\phi)$
is a homogeneous polynomial in $\phi$ of degree $l$ (and likewise
for $S$s and $T$s).

In order that these operators be gauge invariant
we need $l = -3$.  This means that
the first node is trivial, and at the second node the
polynomials are just constants.
At the third node, the polynomials are cubic and
these operators are exactly of the form of marginal deformations
of the superpotential.  Accordingly, they have R-charge $1$.
The image of $Q \dag$ from the R-charge zero operators
at the previous node consists of operators of the form
\eqn\rchargezeroimage{
R_{ab} F^a(\phi) \beta_b \wp, }
where
$R_{ab}$ is a constant and $F^a$ is the
section of $\bigoplus \CO(n_a)$ appearing
in the boundary superpotential.  This
tells us that in $Q \dag$ cohomology,
\eqn\cohorelationonT{
T^a_{(3)}(\phi) \simeq T^a_{(3)} + R_{ab} f^b(\phi).
}
Since there are twenty independent degree three
monomials in four
variables, we find $4(20 - 4) = 64$ elements of cohomology
at R-charge $1$ from this part of the $Q \dag$ complex.

At R-charge $0$, the only operator in $\rm{ ker} (Q \dag)$
is
\eqn\vectorvertex{
 :\beta_a \dag \beta_a: - :\wp \dag \wp: = j_s
}
which by the boundary Gauss' law \gauss\ is
the identity operator which creates (the fermion partner of)
the center-of-mass vector field.

We note here that the brane worldvolume Higgs mechanism has
a very natural implementation in this complex.
Cohomology can appear at adjacent nodes
which have R-charge $0$ and $1$ respectively, resulting
in a new vector field and a new charged multiplet descending
to zero mass.

Note that this framework could as easily have been applied directly
to the large-radius phase of the theory.  In that case, we would
have simply set $p = 0$ to its vacuum value and ignored
its massive fluctuations.  One finds the same operators
representing the cohomology, and the calculation essentially
reduces to the classical mathematics of deformation theory.
It would be interesting to perform an analogous calculation
directly in the Landau-Ginzburg effective field theory, after
integrating out
the $p$ field and gauge multiplet.

\bigskip

\subsubsection{Consequences of the index theorem}

For the Chern-classes of the $SU(3)$
bundle $V$ one finds 
$c_{1}(V)=0$ and $c_{2}(V) = 24$.  A theorem of Mukai \Mukai\ tells
us that the dimension of the moduli space of such a bundle is related
directly to the index of $\Qbar$, and in this example has
(quaternionic) dimension 64.
Therefore, one expects the brane spectrum to include 64 massless 
hypermultiplets,
in agreement with our result.

The index of $\Qbar$ is not directly related to the dimension of
the brane moduli space for branes on Calabi-Yau threefolds.
In those examples, the index of $Q \dag$
is still invariant under smooth deformations of the closed string
and open string moduli, but there are adjacent nodes in the complex
representing scalar states which may pair up with each other while
preserving the index of the $Q\dag$ complex.
Note that the index of $Q\dag$
{\it is} invariant under addition of brane-antibrane pairs
in the linear sigma model.

\newsec{Future Directions}

Many interesting issues arise in the study of D-branes on Calabi-Yau
spaces, and our formalism might be usefully extended to address a number
of them.
Here, we close by mentioning several subjects for future exploration:

\medskip
\noindent
$\bullet$  One expects that generic branes will cross lines of marginal
stability in the $(r,\theta)$ plane.  While a mechanism for
implementing brane decays in the linear model was discussed in
\S5.3, it would be very interesting to derive the form of these loci
of marginal stability directly in the linear model.  This would presumably involve
a direct calculation of the relevant spacetime central charge.

\medskip
\noindent
$\bullet$ The physics of D-branes at
singular points in their moduli space (where even the boundary CFT
becomes ill-defined) should be tractable in this approach.  For closed
strings, the new non-compact branches which arise in the linear model
at singular points in moduli space
were shown in e.g. \evaed\ to allow one to reproduce detailed
calculations about divergent terms in the spacetime effective action.
A similar story may well arise here.
\medskip
\noindent
$\bullet$
A related question: At certain loci in moduli space, wrapped D-branes
become massless spacetime states.  This happens for instance at the
mirror conifold point in the K\"ahler moduli space of the quintic, where
the wrapped D6 brane becomes massless.  What is the behavior of the
worldvolume theory on the brane in such a limit?  What happens as one
makes extremal transitions to new branches of moduli space where such wrapped
branes become fundamental string states?

\medskip
\noindent
$\bullet$ We have confined ourselves to discussing $B$-type branes
in this paper; but similar methods could work for $A$-type branes
as well (for earlier work in this direction see
\refs{\horiiqvafa,\govinda}, for later work in this direction see
\morehell).
As discussed in \refs{\kklmone,\kklmtwo,\agvafa}
there are expected to be intricate, disc-instanton generated
superpotentials for $A$-type branes.  It would be interesting
to formulate a linear sigma model description of such branes in which
the instanton sum was computable.  In a somewhat analogous problem with
more supersymmetry, Morrison and Plesser did succeed in reproducing
closed string instanton effects directly in the linear sigma model
\morrpless.  For recent work in this direction, see
\lastnite.

\medskip
\noindent
$\bullet$ Finally, any microscopically consistent model with wrapped,
space-filling D-branes will have to include orientifolds (or
anti-branes) as well to
cancel RR tadpoles.
It will be interesting to study new phenomena that arise in generalizing this
kind of worldsheet description to models with orientifolds and/or
antibranes.
\medskip

\centerline{\bf{Acknowledgements}}

We are grateful to Jacques Distler,
Kentaro Hori, Greg Moore, Michael Peskin, Joe Polchinski,
Eva Silverstein, Lenny Susskind and Nick Warner for helpful discussions.
Some discussions of boundary conditions
were originally developed
in unpublished work by Thiagarajan Jayaraman,
A.L., Hirosi Ooguri and Tapobrata Sarkar.
They have since appeared
(with the absent authors' blessings) in \govinda.
A.L. would like to thank the other authors and
Suresh Govindarajan for collaboration and discussions.
Parts of this work were undertaken while S.K.
and A.L. enjoyed the hospitality of the
Aspen Center for Physics and the Institute for Theoretical
Physics at Santa Barbara.
This work was supported in part by the DOE under contract
DE-AC03-76SF00515.
The work of S.K. is also supported by a David and Lucile
Packard Foundation Fellowship for Science and Engineering,
an Alfred P. Sloan Foundation Fellowship, and National Science
Foundation grant PHY00-97915.  The work of J.M. 
was supported in part by the Department of Defense 
NDSEG Fellowship program.

\medskip

\appendix{A}{Transformation properties of bulk supermultiplets}

Throughout this paper we use
$Q = {Q_+ - Q_- \over \sqrt 2}$ and its conjugate as the generators
of the unbroken B-type supersymmetry.
$S = {Q_+ + Q_- \over \sqrt 2}$ and its conjugate generate the
supersymmetry transformations broken by the boundary theory.  

The $(2,2)$ algebra is 
\eqn\algebra{
\eqalign{
 \{ Q_\pm , Q_\pm \} = 0 ~~~~&  \{ Q_\pm, Q_\mp \} = 0  \cr
 \{ Q_\pm , Q_\pm^\dagger \} &= 2 P_\pm \cr
  \{ Q_+, Q_-^\dagger \} = 2 Z  ~~~~&   \{ Q_-, Q_+^\dagger \} = 2 Z\dag \cr
 [ Q_\pm, P_\mp ] = - i\sqrt 2 q \lambda_\mp^\dagger ~~~~ &
 [ Q_\pm^\dagger, P_\mp ] = -i \sqrt 2 q \lambda_\mp 
}}
where $Z$ acts by multiplication by $ q \sigma\dag $
on a field of charge $q$ and
$ P_\pm = - i D_\pm $ is the gauge covariant momentum (but does 
not contain $\sigma $s).  Note that 
$ [\nabla_\pm, Q] = [\nabla_\pm, Q\dag] = 0 $.

In the following, our supercharges act by graded commutation.  
The $(2,2)$ supersymmetry transformations of the fields
in a chiral multiplet, $\phi$, of $U(1)$ gauge charge $a$ in
Wess-Zumino gauge are

\eqn\twotwobetter{\eqalign{
	Q \phi = - i \Theta ~~~~ Q\dag \phi = 0 ~~~~~~
	&Q \phi \dag = 0 ~~~~ Q \dag \phi \dag = - i \Theta \dag \cr
	S \phi = - i \eta ~~~~ S \dag \phi = 0 ~~~~~~
	&S \phi \dag = 0 ~~~~ S \dag \phi \dag = - i \eta \dag \cr
	Q \Theta = 0 ~~~~ Q \dag \Theta = 2 \nabla_0 \phi ~~~~~~
	&Q \Theta \dag = 2 \nabla_0 \phi\dag ~~~~ Q \dag \Theta \dag = 0 \cr
	S \Theta = -F ~~~~ S \dag \Theta = 2 \nabla_1 \phi ~~~~~~
	&S \Theta \dag = 2 (\nabla_1 \phi) \dag 
	~~~~ S \dag \Theta \dag = - F \dag \cr
	Q \eta = F ~~~~ Q \dag \eta = 2 \tilde \nabla_1 \phi ~~~~~~
	&Q \eta \dag = 2 (\tilde \nabla_1 \phi)\dag ~~~~ Q \dag \eta \dag = F\dag \cr
	S \eta = 0 ~~~~ S \dag \eta = 2 \tilde \nabla_0 \phi ~~~~~~
	&S \eta \dag = 2 \tilde \nabla_0 \phi \dag 
	~~~~ S \dag \eta \dag = 0  \cr
	Q F = 0 ~~~~~~
	&Q F \dag = 2i\left( (\tilde \nabla_1 \Theta) \dag -
	\nabla_0 \eta \dag 
	\right)
	+ 2 q \phi \dag \left( \lambda_+ + \lambda_- \right) \cr
	Q \dag F = 2i\left(\tilde \nabla_1 \Theta -
	\nabla_0 \eta \right)
	+ 2 q \phi\left( \lambda_+\dag + \lambda_- \dag \right) 
	 ~~~~~~
	&Q \dag F \dag = 0 \cr
	S F = 0 ~~~~~~
	&S F \dag = 2i\left( \tilde \nabla_0 \Theta\dag  -
	(\nabla_1 \eta) \dag \right)
	+ 2 q \phi \dag \left( \lambda_+ - \lambda_-  \right) \cr
	S \dag F = 2i \left( \tilde \nabla_0 \Theta -
	\nabla_1 \eta \right)
	+ 2 q \phi\left( \lambda_+\dag - \lambda_- \dag \right)
	~~~~~~
	&S \dag \Theta \dag = - F \dag
}}

The vector multiplet fields transform as:
\eqn\twotwovector{
\eqalign{
	Q \sigma = i\lambda_+\dag ~~~~ Q \dag \sigma = - i \lambda_- ~~~~~~
	&Q \sigma \dag = - i \lambda_-\dag ~~~~ Q \dag \sigma \dag = i \lambda_+ \cr
	S \sigma = i \lambda_+\dag  ~~~~ S \dag \sigma = i \lambda_- ~~~~~~
	&S \sigma \dag = i \lambda_- \dag ~~~~ S \dag \sigma \dag =  i \lambda_+ \cr
	Q \lambda_+ = - \del_+ \sigma \dag - {1 \over \sqrt 2 }
	\left (i D - {v_{+-} \over \sqrt 2} \right)~~~~~~
	&Q \dag \lambda_+ \dag =
	- \del_+ \sigma  + {1 \over \sqrt 2 } \left(i D + {v_{+-} \over \sqrt 2}\right) \cr
	Q \dag \lambda_+ = 0 ~~~~ S \dag \lambda_+ = 0 ~~~~~~
	&Q \lambda_+ \dag = 0 ~~~~ S \lambda_+ \dag = 0 \cr
	S \lambda_+ = - \del_+ \sigma \dag + {1 \over \sqrt 2 }
	\left (i D - {v_{+-} \over \sqrt 2}\right)~~~~~~
	&S \dag \lambda_+ \dag =
	- \del_+ \sigma  - {1 \over \sqrt 2 } \left(i D + {v_{+-} \over \sqrt 2}\right) \cr
	Q \lambda_- =  \del_- \sigma + {1 \over \sqrt 2 }
	\left (i D + {v_{+-} \over \sqrt 2} \right)
	~~~~~~&Q \dag \lambda_- \dag =
	\del_- \sigma\dag  - {1 \over \sqrt 2 } \left(i D - {v_{+-} \over \sqrt 2}\right) \cr
	Q \dag \lambda_- = 0 ~~~~
        S \dag \lambda_- = 0 	~~~~~~
	&Q \lambda_- \dag = 0 ~~~~
	S \lambda_- \dag = 0  \cr
	S \lambda_- = - \del_- \sigma  + {1 \over \sqrt 2 }
	\left (i D + {v_{+-} \over \sqrt 2}\right)
	~~~~~~&S \dag \lambda_- \dag =
	- \del_- \sigma \dag - {1 \over \sqrt 2 } \left(i D - {v_{+-} \over \sqrt 2}\right) \cr
	Q v_{+-} = - i \left( \del_+ \lambda_- \dag + \del_- \lambda_+\dag \right) ~~~~~~
	&Q \dag v_{+-} =  - i \left( \del_+ \lambda_- + \del_- \lambda_+ \right) \cr
	S v_{+-}  = - i \left(  \del_+ \lambda_- \dag - \del_- \lambda_+\dag \right) ~~~~~~
	&S \dag v_{+-}  = - i \left( \del_+ \lambda_-  - \del_- \lambda_+ \right) \cr
	Q i\sqrt 2 D = i \left( \del_+ \lambda_- \dag - \del_- \lambda_+\dag \right)~~~~~~
	&Q \dag i\sqrt 2 D =  - i\left( \del_+ \lambda_- - \del_- \lambda_+ \right) \cr
	S i\sqrt 2 D  = i \left(  \del_+ \lambda_- \dag + \del_- \lambda_+\dag \right) ~~~~~~
	&S \dag i\sqrt 2 D  =  -i \left( - \del_+ \lambda_-  + \del_- \lambda_+ \right)
}
}

\appendix{B}{Boundary superspace}

We can exponentiate the supersymmetry transformations 
to define a superspace formalism.  
We use the same bulk superspace coordinates as in \phases.
In this language the action of supersymmetry on superfields is generated
by:
\eqn\superderiv{
\eqalign{
	\delta \Phi & = \left(\epsilon^\alpha Q_\alpha
		- \bep^\alpha \Qbar_\alpha \right)\Phi\cr
	&= \left( -\ep_- Q_+ + \ep_+ Q_- + \bep_- \Qbar_+
		- \bep_+ \Qbar_- \right) \Phi
}}
where $\epsilon^\alpha$ is the Grassman parameter; the spinor indices
are raised and lowered by the antisymmetric tensor
$\epsilon_{\alpha\beta}$ as in \phases.  Recall that
\eqn\qdef{
\eqalign{
	Q_+ & = \frac{\del}{\del\theta^+} + i \bthet^+ \p_+\cr
	Q_- & = \frac{\del}{\del\theta^-} + i \bthet^- \p_-\cr
	\Qbar_+&=-\frac{\del}{\del\bthet^+}-i\theta^+\p_+\cr
	\Qbar_-&=-\frac{\del}{\del\bthet^-}-i\theta^-\del_-
}}
where
$$ \del_\pm = \del_0 \pm \del_1 $$
We can similarly define superspace derivatives:
\eqn\ddef{
\eqalign{
	D_\pm &= \frac{\del}{\del\theta^\pm}-i\bthet^\pm\del_\pm\cr
	\Dbar_\pm &=-\frac{\del}{\del\bthet^\pm}+i\theta^\pm\del_\pm\ .
}}

We wish to preserve as symmetries 
those transformations for which $\ep_+ = \ep_-$.  Define:
\eqn\newparam{
\eqalign{
	\ep=\frac{1}{\sqrt{2}}\left(\ep_+ +\ep_-\right)\cr
	\tilde{\ep}=\frac{1}{\sqrt{2}}\left(\ep_+ -\ep_-\right)
}}
and set $\tilde{\ep} = 0$.
Then
\eqn\newvariat{
	\delta\Phi = \left( \ep \frac{Q_- - Q_+}{\sqrt{2}}
	+ \bep \frac{\Qbar_+ - \Qbar_-}{\sqrt{2}} \right) \Phi
}
We can define
\eqn\boundgrass{
	\theta = \frac{\theta^+ - \theta^-}{\sqrt{2}}
}
and
\eqn\boundsusy{
\eqalign{
	Q&=\frac{Q_+ -Q_-}{\sqrt{2}}=\frac{\del}{\del\theta}
	+i\bthet\del_0 ~~~~
	\Qbar=\frac{\Qbar_+ - \Qbar_-}{\sqrt{2}}=
	-\frac{\del}{\del\bar{\theta}}-i\theta\bp_0\ ,
}}
so that
$$\delta \Phi = \left(\ep Q-\bep \Qbar\right)\Phi\ .
$$
Similarly, we can define superspace derivatives:
\eqn\boundD{
\eqalign{
	D&=\frac{D_+-D_-}{\sqrt{2}}=\frac{\del}{\del\theta}
		-i\bthet\del_0\cr
	\Dbar&=\frac{\Dbar_+-\Dbar_-}{\sqrt{2}}=
		-\frac{\del}{\del\bthet}+i\theta\del_0\ .
}}

\appendix{C}{The tensor formula}

Here we state a formula for the tensor product
of two sheaves.  The
support of this product sheaf is the intersection of the
supports of the original sheaves.

Take a complex
\eqn\sequenceV{
0 \to V_0 {\buildrel d_1 \over\longrightarrow} V_1
	{\buildrel d_2 \over\longrightarrow} 
	\cdots {\buildrel d_m\over\longrightarrow} V_m \to 0
}
whose cohomology
is a sheaf $E$, and a complex
\eqn\sequenceW{ 
	0 \to W_0 {\buildrel \tilde d_1 \over\longrightarrow} W_1
	{\buildrel \tilde d_2 \over\longrightarrow} \cdots
	{\buildrel \tilde d_n \over\longrightarrow} W_n \to 0
}
whose cohomology is a sheaf $F$. 
Then a complex whose cohomology is $E \otimes F$ is:
\eqn\tensorformula{
0 \to V_0 \otimes W_0 {\buildrel D_1 \over\longrightarrow}
V_0 \otimes W_1 \oplus V_1 \otimes W_0
{\buildrel D_2 \over\longrightarrow}
V_0 \otimes W_2 \oplus V_1 \otimes W_1 \oplus V_2 \otimes W_0
{\buildrel d_3\over\longrightarrow} \cdots \to 0\ ,
}
where
\eqn\derivation{
D_{r+s+1} (a \otimes b)
= d_{r+1}(a) \otimes b + (-1)^r a \otimes \tilde d_{s+1}(b)
}
for $a \in V_r$ and $b \in W_s$.

This formula is well known to mathematicians, to whom, however,
the minus signs in \derivation\ seem mysterious
(\cf\ p. 431 of \eisenbud).
Rather than proving this directly as a mathematical statement, we can give
a physical interpretation.
Consider two sets of fiber fields $\b,\wp$ and
$\bt,\wpt$ which transform under two \it independent \rm boundary symmetries
$U(1)\ll{1,2}$.
The $\b,\wp$ fields are linear model fields satisfying the correct deformed
chiral constraints to have cohomology equal to $E$.  That is
\eqn\qonnontilded{
\eqalign{
\{\Qbar , \b_{2n} \} & = d_{2n-1} \wp\dag_{2n-1}
\cr
[ \Qbar, \wp\dag_{2n+1} ] & =  d_{2n} \b_{2n}\ .
}}
Similarly, $\bt,\wpt$ satisfy the contraints:
\eqn\qontilded{
\eqalign{
\{\Qbar , \bt_{2n} \} & = \tilde{d}_{2n-1} \wpt\dag_{2n-1}
\cr
[ \Qbar, \wpt\dag_{2n+1} ] & =  \tilde{d}_{2n} \bt_{2n}
}}
in order to describe $F$.  We can combine $\beta_k,\wp\dag_k$
into an object $\gamma_m$ which is even(odd) when $m$ is
odd(even). 
If the left endpoint lives in $E\otimes F$, the fields
for {\it both} $E$ and $F$ should be excited.
Thus we perform two boundary charge projections, 
each onto charge sector $+1$.

To see that this is equivalent to the complex
\tensorformula\ with exterior
derivative $D$, examine variables invariant 
under the combination $U(1)_1 - U(1)_2$.
The gauge invariant combinations which satisfy deformed chiral constraints
off shell are
$\b\bt, \b\wpt\dag, \wp \dag \bt,\wp \dag \wpt \dag$.
These combinations all have charge $+1$ under the remaining
symmetry ${1\over 2} (U(1)_1 + U(1)_2)$, and satisfy:
\eqn\cpderivation{
[\Qbar, \g_m \tilde{\g}_n] = d_{m-1} (\g_{m-1} ) \tilde{\g}_n 
	+ (-1)^{m-1}\g_{m}\tilde{d}_{n-1}(\tilde{\g}_{n-1})\ .
}
Thus, the singlets of the confined gauge group $U(1)_1 - U(1)_2$
transform under supersymmetry as fibers 
in the complex defining $E\otimes F$.

\listrefs 
\end